\documentclass{aastex61}
\usepackage{amsmath,amstext}
\usepackage[T1]{fontenc}
\usepackage{apjfonts} 
\usepackage[figure,figure*]{hypcap}
\usepackage{booktabs}
\usepackage{multirow}
\usepackage{makecell}

\usepackage{color, xcolor}	
\usepackage{comment}

\submitjournal{ApJL}

\shorttitle{Dynamical TDE unification}
\shortauthors{Thomsen et al.}

\begin{document}

\title{Dynamical Unification of Tidal Disruption Events}
\correspondingauthor{Lixin Dai lixindai@hku.hk}

\author[0000-0003-4256-7059]{Lars Lund Thomsen}
\affiliation{Department of Physics, University of Hong Kong, Pokfulam Road, Hong Kong, China}

\author[0000-0003-0509-2541]{Tom Kwan}
\affiliation{Department of Physics, University of Hong Kong, Pokfulam Road, Hong Kong, China}

\author[0000-0002-9589-5235]{Lixin Dai}
\affiliation{Department of Physics, University of Hong Kong, Pokfulam Road, Hong Kong, China}

\author[0000-0003-2872-5153]{Samantha Wu}
\affiliation{California Institute of Technology, Astronomy Department, Pasadena, CA 91125, USA}

%\author[0000-0002-6485-2259]{Nathaniel Roth}
%\affiliation{Lawrence Livermore National Laboratory, Livermore, CA 94550, USA}

\author[0000-0003-2558-3102]{Enrico Ramirez-Ruiz}
\affiliation{Department of Astronomy and Astrophysics, University of California Santa Cruz, 1156 High Street, Santa Cruz, CA 95060, USA}
\affiliation{DARK, Niels Bohr Institute, University of Copenhagen, Denmark} %\\

\begin{abstract}
About a hundred tidal disruption events (TDEs) have been observed and they exhibit  a wide range of emission properties both at peak and over their lifetimes. Some TDEs peak predominantly at X-ray energies while others radiate chiefly at UV and optical wavelengths. While the peak luminosities across TDEs show distinct  properties,  the evolutionary behavior can also vary between TDEs with similar peak emission properties. At late time, some optical TDEs rebrighten in X-rays, while others maintain strong UV/optical emission components. In this Letter, we conduct three-dimensional general relativistic radiation magnetohydrodynamics
simulations of TDE accretion disks at varying accretion rates ranging from a few to a few tens of the Eddington accretion rate. We make use of Monte Carlo radiative transfer simulations to calculate the reprocessed spectra at various inclinations and at different evolutionary stages. We confirm the unified model proposed by \citet{Dai18}, which predicts that the observed emission largely depends on the viewing angle of the observer with respect to the disk orientation (X-ray strong when viewed face-on and UV/optically strong when viewed disk-on). What is more, we find that disks with higher accretion rates have  elevated wind and disk densities, which increases the reprocessing of the high-energy radiation and thus generally augments the optical-to-X-ray flux ratio along a particular viewing angle. This implies that at later times, as the accretion level declines, we expect the funnel to effectively open up and allow more X-rays to leak out along intermediate viewing angles. Such dynamical model for TDEs can provide a natural explanation for the diversity in the emission properties observed in TDEs at peak and along their temporal evolution.
\end{abstract}

\section{Introduction}
\label{sec:intro}

\noindent The tidal disruption of stars by massive black holes (MBHs) offers a unique probe of MBH demographics \citep{Mockler19, Gezari21review}, host galaxy properties \citep{French20}, stellar dynamics \citep{Stone20review, Pfister20}, as well as black hole accretion and jet physics \citep{Dai21review}. When a star with mass $m_\star$ and radius $r_\star$ approaches a MBH with mass $M_{\rm BH}$, the star is disrupted within the tidal radius $R_t \approx (M_{\rm BH}/ m_\star)^{1/3} r_\star$, where the MBH's tidal force exceeds the stellar self-gravity. About half stellar debris orbits back to the vicinity of the MBH following a characteristic pattern which first quickly increases to a peak and then declines with time following a mass fallback rate $\dot{M}_{\rm fb} \approx  t^{-5/3}$ \citep{Rees88, Evans89, Guillochon13}. When $M_{\rm BH}\sim 10^6 M_\odot$, $\dot{M}_{\rm fb}$ can exceed the Eddington accretion rate by two orders of magnitude at peak and stays super-Eddington for years after peak. Here we define the Eddington accretion rate of a black hole as $\dot{M}_{\rm Edd} = L_{\rm Edd}/(\eta_{\rm NT}\ c^2)$,
where $L_{\rm edd}=4\pi GM_{\rm BH}c/\kappa$ is the Eddington luminosity for an opacity $\kappa$, and $\eta_{\rm NT}$ is the nominal accretion efficiency for the Novikov-Thorne thin disk solution \citep{Novikov73}.

With the recent launches of all-sky transient surveys including ZTF and eROSITA, the number of detected TDE candidates has reached around one hundred \citep[e.g.,][]{Gezari21review, Sazono21, Hammerstein22}. TDEs have been categorized into two classes based on their main emission type at peak: X-ray TDEs \citep{Auchettl17,Saxton21review} and optical TDEs \citep{vanVelzen20review}. Most X-ray-selected TDEs emit soft X-rays  with effective temperatures at $10^5-10^6$K, while only three of them emit beamed, hard X-rays which are associated with relativistic jets \citep[e.g.,][]{Bloom11, Burrows11, Cenko12, DeColle12}. The optical TDEs have lower effective temperatures at few$\times10^4$K, and are further characterized by their spectroscopic features (producing strong and broad H, or He, or Bowen fluorescence emission lines) \citep{Leloudas19, Charalampopoulos22}. Interestingly, as the luminosity of a TDE typically declines by around one order of magnitude over tens to hundreds of days after peak, its effective temperature usually undergoes a peculiar non-evolution. Recently, a large number of TDEs have been followed up for longer than a year and they show different late-time behaviors.  A few initially optically-strong TDEs rebrighten in X-rays \citep[e.g.,][]{Gezari17, Holoien18, Wevers19, Hinkle21, Liu22}, while many others maintain strong UV/optical emissions over years \citep{vanVelzen19}. 

Many theoretical models have been proposed to explain these TDE emission properties \citep{Roth20review}. While the X-ray emissions have been predicted by classical accretion disk models \citep{Ulmer99}, UV/optical emissions are argued to be produced from either the reprocessing of X-ray photons in an extended envelope or outflows \citep{Loeb97, Strubbe09, Lodato11, Coughlin14, Guillochon14, Metzger16, Roth16} or the shocks powered by debris stream self-intersection \citep{Piran15, Bonnerot21}.
The late-time rebrightening of X-rays in TDEs can be accounted by either a change in the disk morphology as the accretion rate declines from super-Eddington to sub-Eddington, or the delayed onset of accretion \citep{vanVelzen19, Hayasaki21}. The latter model, however, is disfavored by recent simulations which show that a large fraction of the debris already settles into a disk with moderate eccentricity within dynamical times for at least a subset of TDE parameters \citep{Bonnerot21, Andalman22}.

In search for a unified model that can explain both the optical and X-ray TDEs, \citet{Dai18} have carried out the first three-dimensional (3D) general relativistic radiation magnetohydrodynamics (GRRMHD) simulation tailored for TDE super-Eddignton accretion flow. The simulated disk, around a black hole with $M_{\rm BH} = 5\times10^6 M_\odot$ and spin parameter $a=0.8$, has an accretion rate of $\dot{M}_{\rm acc} \sim 10 \ \dot{M}_{\rm Edd}$, representing a typical accretion level in TDEs. The spectra of the disk have been obtained by post-processing the simulated disk with a novel radiative transfer code. It is found that the observed emission type largely depends on the viewing angle of the observer with respect to the disk orientation. When viewed face-on, X-ray emissions can escape from the optically-thin funnel surrounded by winds. When viewed edge-on, X-ray emissions are heavily reprocessed in the geometrically and optically thick wind and disk, so only UV/optical emissions can be observed. 

While this study gives a good first-order description of TDE disks and spectra, recent simulations also show that the properties of a super-Eddington disk depends on the accretion rate, the black hole mass and spin, as well as the magnetic flux threading the disk \citep{Jiang14, Jiang19, McKinney15, Sadowski16}. Although the black hole spin and disk magnetic flux might only mildly affect the main structure of the disk, they determine whether a relativistic jet can be launched \citep{Blandford77, Curd19}.  At first glance, the black hole mass is not expected to affect the disk spectra significantly, since most TDE host MBHs have masses in a narrow range $\approx 10^5-10^7\ M_\odot$. However, the peak fallback rate of the TDE debris sensitively depends on the black hole mass with the relation $\dot{M}_{\rm fb, peak}/\dot{M}_{\rm Edd}\propto M^{-3/2}_{\rm BH}$. Therefore, TDEs from smaller MBHs should in general have much higher accretion rates at peak than those from larger MBHs \citep{RamirezRuiz09}. Also, further variance of the accretion rates in different TDEs as $\dot{M}_{\rm fb, peak}$ can be introduced by the difference in the masses and ages of the disrupted stars \citep{Law-Smith20}.

In this Letter, we investigate how the TDE disk structure and the accompanied emission are influenced by the accretion rate at super-Eddington rates. We conduct three 3D GRRMHD simulations of super-Eddington disks with similar structures but varying accretion rates, and post-process the simulated disks to obtain their spectra at different inclination angles. These simulations are useful for understanding the diversity of the emissions observed from different TDEs, as well as the evolution of single TDEs as their fallback and accretion rate decline after peak.  In Section \ref{sec:method} we introduce the setup of the GRRMHD and radiative transfer simulations. In Section \ref{sec:results} we give the main results and compare with TDE key observables. In Section \ref{sec:summary} we draw a summary and discuss the implications and future work.

\section{Methodology}
\label{sec:method}

\subsection{Disk simulation setup}
\label{sec:GRRMHDsetup}
\noindent 
We carry out three dimensional simulations of super-Eddington disks using a fully 3D GRRMHD code {\tt HARMRAD} with M1 closure \citep{McKinney14}. In all simulations the MBH has $M_{\rm BH}=10^6 M_{\odot}$ and $a=0.8$. 
The radial grid has 128 cells spanning from  $R_{\rm{} in}\approx 1.2 R_g $ to $R_{\rm out}=8500 R_g$ ($R_g=GM_{\rm BH}/c^2$ is the gravitational radius of the MBH), with cell size increasing exponentially until $R_{\rm break} \approx 500 R_g$ and then increasing hyper-exponentially. The $\theta$-grid has 64 cells spanning from 0 to $\pi$ with finer resolution in the jet and disk regions. The $\phi$-grid has 32 grids spanning uniformly from 0 to $2\pi$ with periodic boundary conditions. The gas is assumed to have solar chemical abundances (mass fractions of H, He, and ``metals'', respectively, $X = 0.7, Y = 0.28, Z = 0.02$). Frequency-mean absorption and emission opacities are used as in \citet{McKinney15}, except that the Chianti opacity is removed as it is unimportant for the disk temperature of our TDE models. Thermal Comptonization is also included. 

We tailor the disk initial conditions to be consistent with realistic TDE scenarios following the setup described in \citet{Dai18}. The accretion disk is initialized with Keplerian velocity profile with a rest-mass density which is Gaussian in angle with a height-to-radius ratio of $H/R\approx 0.3$, so that the initial density profile is given by
\begin{equation}
    \rho(r,z) = \rho_0 r^{-1.3}  e^{-z^2/(2H^2)},
\end{equation}
where $H$ is the scale height, $\rho_0$ is the initial reference density. The disks are checked to have total masses and specific angular momentum consistent to the first order with the conditions in TDEs. Adjusting $\rho_0$ leads to different accretion rates after the disk reaches the quasi-equilibrium state. As $\dot{M}_{\rm fb, peak}$ can reach $\approx 100 \dot{M}_{\rm Edd}$ in a TDE around a $10^6 M_\odot$ black hole, and a large fraction of the debris mass is expected to escape in outflows, we set the aimed accretion rates to be a few to a few tens of $\dot{M}_{\rm Edd}$ (Table \ref{tab:quantities}).

\begin{comment}
The lower resolution vertically across the disk at large radii shall lead to unstable dynamics evolution, so we have truncated the disk at $R_{t}=500 R_g$ to prevent gas from the regions of large radii contaminating the evolution of the inner accretion disk. To do so, we set an exponential decay such that 
\begin{equation}
    \rho(r\geq R_t) = \rho(r) e^{-r/R_t+1}
\end{equation}

The total ideal pressure $P_{\rm{}tot}=(\Gamma_{\rm{}tot}-1)u_{\rm{}tot}$ is initially randomly perturbed by 10 percent to seed the MRI, $\Gamma_{\rm{}tot}=4/3$, $u_{\rm{}tot}$ is the total internal energy density. The disc gas has $\Gamma_{\rm{}gas}=5/3$. The disc is surrounded by an initial atmosphere with rest-mass density $\rho=10^{-5}(r/R_g)^{-1.1}$ and gas internal gas density $e_{\rm{}gas}=10^{-6}(r/R_g)^{-5/2}$. The disc radiation energy density and flux are set by local thermal equilibrium (LTE) and flux-limited diffusion \citep{McKinney14}, radiation atmosphere is neglected.
\end{comment}

A large-scale poloidal magnetic field is initially seeded. As adopted in \citet{McKinney15}, for $r$ smaller than a breaking radius $R_b = 500R_g$, the $\phi$ component of the vector potential is given by
$A_{\phi}=MAX(r \ 10^{40}-0.02, 0)(\sin\theta)^5$. For $r \geq R_b$, the field transitions to a split monopolar field, which is given by
$A_{\phi}=MAX(R_b \ 10^{40}-0.02, 0)(\sin\theta)^{1+4(R_b/r)}$. 
The field is normalised so that the initial ratio of gas+radiation pressure to magnetic pressure $\beta \equiv (p_{\rm{}gas}+p_{\rm{}rad})/p_{b}$ has a minimum value of 10.

\subsection{Radiative transfer setup}
\label{sec:RTsetup}

\noindent 
We use the Monte Carlo radiative transfer code {\tt SEDONA} \citep{Kasen06} to post-process the simulated disks and calculate the escaping radiation. More specifically, we calculate the bound electron level populations under non-local thermal equilibrium (LTE) as in \citet{Roth16} and the Comptonization of electrons as in \citet{Roth17}. We track the free-free interactions, as well as the bound-bound and bound-free interactions with their opacities obtained from. the atomic database {\tt CMFGEN}. The gas is assumed to consist of only H, He, and O with solar abundances. We focus on calculating the SED, and leave the line modeling to a future study which requires higher-solution simulations. 

For each accretion rate, we calculate the spectra for four inclination bins of $\theta_{\rm bin}=$ 10, 30, 50 and $70^\circ$. In each bin, we take an average over $\theta_{\rm bin} \pm 5^\circ$ for the already time-and-$\phi-$averaged profile of the simulated disk. Since the simulated jet density is likely numerically boosted, for the bins at $\theta_{\rm bin}=$ 10 and 30$^\circ$, we also reduce the jet density by 100 times before taking the average. We note that the first-order behavior of the spectra does not depend on this arbitrary choice of density re-scaling inside the jet. 
We set the source to be a blackbody spectrum with a single temperature of $10^6$K, which is consistent with the inner disk temperature. The source photons are injected from an inner boundary set at the boundary between the inflow and outflow, which is typically at a few $R_g$ except for the bin at $i=70^\circ$ which is partly in the disk inflow region. For the latter, we place the inner boundary at $R=20R_g$ instead, and set the gas velocity to be always zero.
The photons are then propagated in 3D, by assuming that the gas density, temperature, and radial velocity profiles are spherically symmetric. The photons propagate outwards until they leave the computational domain set at $R=4000R_g$ or have become a part of the kinetic/thermal pool. 

Based on the Monte Carlo calculations, we iterate the gas temperatures, gas ionization state, bound electron states, and radiative transfer solution under the assumption of radiative equilibrium until a steady solution has been reached (after approximately 20 iterations). Since GRRMHD simulations show that the luminosity of super-Eddington disks are not always capped by $L_{\rm Edd}$, we also tweak the source photon luminosity and obtain the spectra under two limits -- the escaped bolometric  luminosity is either $L_{\rm bol}=L_{\rm Edd}\approx 10^{44} \rm \ erg s^{-1}$ or $L_{\rm bol}= 0.1 \ \dot{M}_{\rm acc} c^2 \approx (\dot{M}_{\rm acc} / \dot{M}_{\rm Edd})\ L_{\rm Edd}$.

\section{Results}
\label{sec:results}

\subsection{Properties of the accretion flow} \label{sec:results:accretionflow}

\noindent In all three GRRMHD simulations, it takes an initial time period of $t= {\rm few} \times1,000 R_g/c$ for the accretion flow to get established. As the disk evolves, magnetic fluxes accumulate near the MBH and their strength further grow via the magnetorotational instability (MRI) \citep{Balbus98}.
Here we focus on the final stage of the simulations when the accretion flow has reached steady state and the wind has established equilibrium at most inclination angles. The black hole parameters and some basic quantities of the accretion disks, averaged over the period of $t= 15,000-20,000 R_g/c$, are listed in Table \ref{tab:quantities}. The disk profiles used for post-processing are  also first $\phi-$averaged and then time-averaged over the same period. More numerical details  of the simulated disks are given in Appendix \ref{app:simulation}.

\begin{deluxetable*}{cccccc}[h]
\tablecaption{Black hole and accretion disk parameters  \label{tab:quantities}}
\tablewidth{0pt}
\tablehead{
\colhead{Model} &   \colhead{$M_{\rm{}BH}\ (M_{\odot})$}  & \colhead{$a$}  & \colhead{$\dot{M}_{\rm acc} \ (\dot{M}_{\rm{}Edd})$} & \colhead{$\dot{M}_w\ [\dot{M}_{\rm{}Edd}]$}  & \colhead{$L_{\rm{}RAD}\ (L_{\rm{}Edd})$}
}
\startdata
M6a08-1	&  $10^6$   &   0.8 &   7   &  1.4	&	5.4   \\
M6a08-2	&  $10^6$   &   0.8 &   12  &  4.5	&	3.3   \\
M6a08-3	&  $10^6$   &   0.8 &   24  &  14	&	8.1   \\
\enddata
\end{deluxetable*}

\begin{figure}[h]
    \centering
    \includegraphics[width=0.95\textwidth]{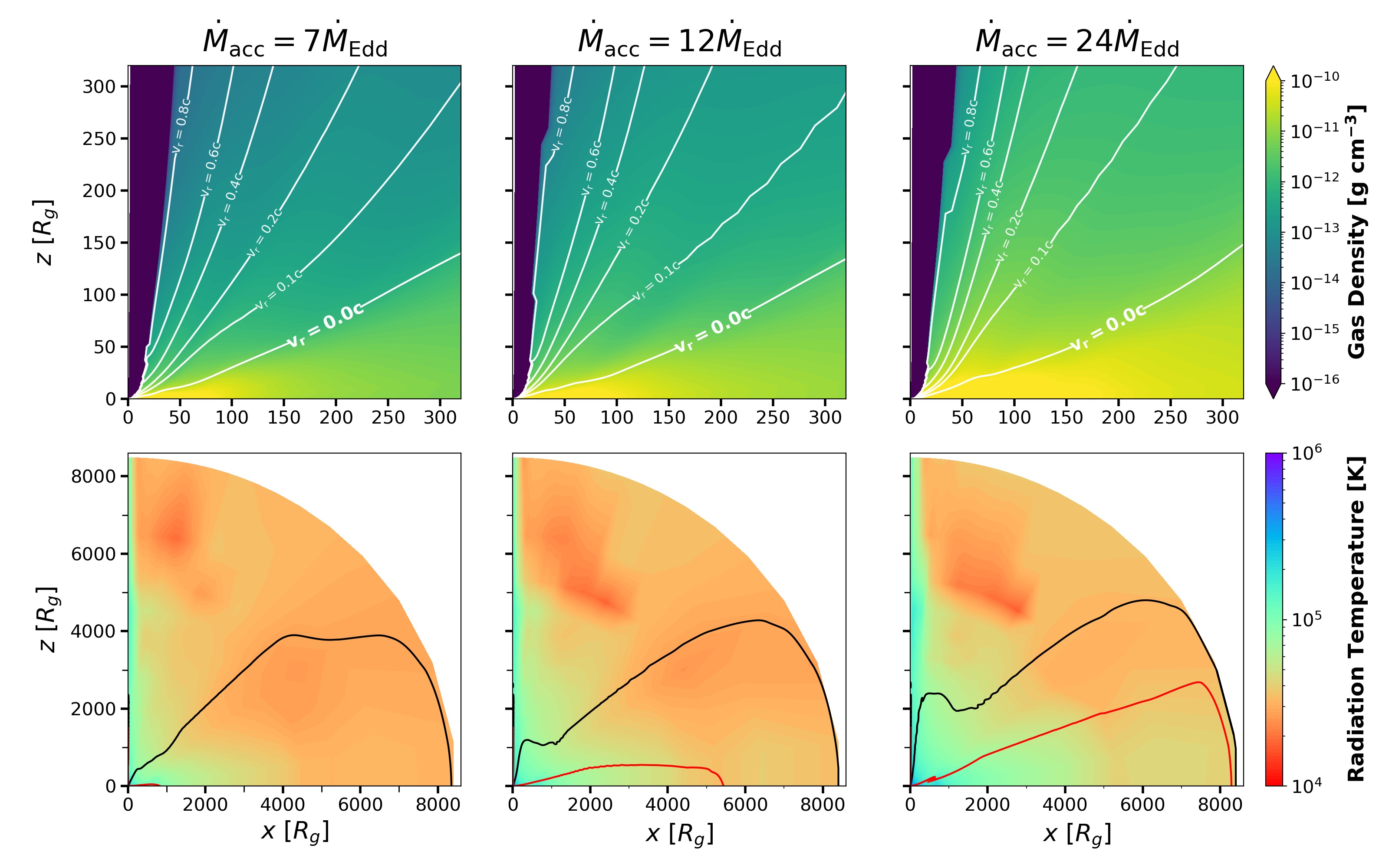}
    \caption{The 2D vertical profiles of time-and-$\phi$-averaged gas rest-mass density $\rho_0$ (upper panels, zoomed into the inner regions) and radiation temperature $T_{\rm{}RAD}$ (lower panels, whole range of the simulation box), for the three runs with different accretion rates (from left to right: $\dot{M}_{\rm acc}=7\dot{M}_{\rm Edd}$, $12\dot{M}_{\rm Edd}$ and $24\dot{M}_{\rm Edd}$) in the quasi-equilibrium state. In the upper panels, we show the contours of constant lab-frame radial velocity ($v_r\equiv u^r/u^t$)  with white lines, and mark the jet regions where the electromagnetic energy is larger than the rest-mass energy of the gas with dark blue color. In the lower panels, the black lines indicate the electron-scattering photosphere with $\tau_{\rm es}=1$, and the red lines indicate the effective photosphere with $\tau_{\rm eff}=1$. Larger accretion rates induce larger disk/wind density and higher gas/radiation temperature, while the gas distribution and velocity structure remain rather robust against the variance in accretion rates. The sizes of photospheres generally increase as accretion rate increases. }
    \label{fig:GRRMHD}
\end{figure}

The averaged accretion rates onto the event horizon of the MBH in these three simulations are $\dot{M}_{\rm acc}\approx$ 7, 12 and 24 $\dot{M}_{\rm Edd}$. To first order, these super-Eddington accretion flows have similar structures as those described in \citet{Dai18}. 
As shown in the upper panels of Fig. \ref{fig:GRRMHD}, geometrically and optically thick disks are formed with half-angular thickness $H/R \approx 0.20 - 0.25$, with, as expected, higher gas density towards the equatorial plane. The rotation profile of the resultant thick disks is sub-Keplerian.
Wide-angle winds are launched by the large radiation  and magnetic pressure, which are also optically thick at most angles except close to the polar axis. There are roughly two components of the wind: an ultra-fast outflow (UFO) with speeds faster than $0.1c$ within $\lesssim 45^\circ$ from the pole, and a slower but denser wind outside $\gtrsim 45^\circ$ inclination. At larger accretion rates, the disk and wind densities increase, and the winds become  slower as a result of mass-loading.  However, the wind geometry stays fairly similar.

The emission properties of the accretion flow should be examined near the photosphere. The optical depth for an opacity $\kappa$ is calculated using $\tau(r) = \int \rho \kappa \text{d}l$
along the radial direction, $r$, from the outer boundary $R_{\rm{} out}=8500 R_g$ towards the event horizon. (Here relativistic effects are included so that: $\text{d}l = -f_{\gamma} \text{d}r$, with $f_{\gamma} \approx u^{t}(1-v/c)$ and $v/c \approx 1-1/(u^t)^2$).
The electron scattering photosphere is defined by $\tau_{\rm es} (r) =1$ with $\kappa=\kappa_{\rm{}es} \approx 0.2(1+X)~{\rm cm^2~g^{-1}}$. 
The effective photosphere is then defined by $\tau_{\rm eff} (r) =1$ with $\kappa=\kappa_{\rm{}eff}=\sqrt{3\kappa_{\rm{}ff}(\kappa_{\rm{}ff}+\kappa_{\rm{}es})}$, where we have only considered free-free opacity $\kappa_{\rm{}ff} \approx 3.82\times 10^{22} Z(1+X)(1-Z)\rho T_g^{-3.5}$ in the scattering-dominated gas, where $T_g$ is the gas temperature. Both photospheres are plotted over the disk profiles in the lower panels of Fig. \ref{fig:GRRMHD}. As expected, we see the presence of  an optically thin ``funnel'' near the rotation axis where the wind density is correspondingly lower. As the accretion rate increases, the wind becomes more opaque, which reduces  the angular size of the funnel. 

The effective photospheres reside mostly within $r\sim 5000 R_g$ except along the equitorial direction for the disk with the largest $\dot{M}_{\rm acc}$. Therefore, we select to evaluate various physical quantities at $r= 5500 R_g$. This gives averaged wind mass rates of $\dot{M}_w=$ 1.4, 4.5 and 14 $\dot{M}_{\rm Edd}$, and averaged bolometric luminosities at $L_{\rm RAD}$= 5.4, 3.3 and 8.1 $L_{\rm Edd}$, respectively, for the three simulations. 
The radiation temperature of the accretion flow is also plotted, which varies from $\approx 10^6$ K near the black hole to $\approx 10^4$ K in the outer part of the disk and wind. The radiation flux varies with the inclination as shown in \citet{Dai18} and leaks out through the region of least resistance, which is the funnel. 

Magnetic fluxes are dragged by the accretion flow and  accumulate near the MBH. This  brings the inner regions to swiftly become a magnetically arrested accretion disk (MAD) \citep{Narayan03}. Relativistic jets are  launched magnetically through the Blandford-Znajek mechanism \citep{Blandford77} in all simulations. In this Letter we focus on calculating the emission properties from the disk, and leave the analysis of the jets for future study.

\subsection{Spectra from post-processing}

\noindent In this Section we investigate how the emission reprocessing depends on a few key parameters, namely, the viewing angle, the accretion rate, and the luminosity.
The dependence on viewing angle has been previously studied by \citet{Dai18}. They  show that for a fixed accretion rate, there is a clear trend of the spectral energy distribution (SED) moving towards larger wavelengths with increasing inclination angle. At low inclinations, the gas density is lower, so X-rays produced from the inner disk easily escape. At relatively large inclinations, on the other hand, the optically thick wind and outer disk serve as an effective  reprocessing envelope. More specifically, in the fast wind region, the photons lose energy as they go through multiple scatterings in the expanding outflow before escaping. In the disk and the slow wind region, the reprocessing is mainly caused by the absorption of X-ray photons and the thermal Comptonization of electrons.

\begin{figure}[h!]
    \centering
    \includegraphics[width=0.44\textwidth]{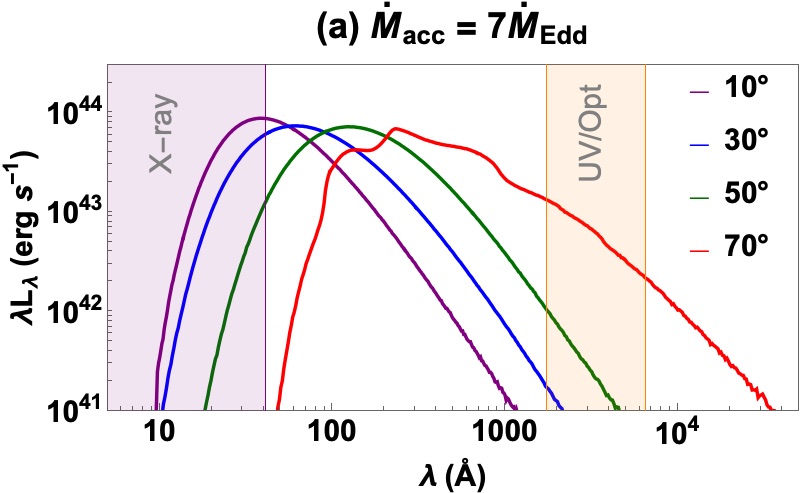}
    \figurenum{2a}
    \label{fig:spectra:acc7}
    \hspace{0.1cm}
        \includegraphics[width=0.44\textwidth]{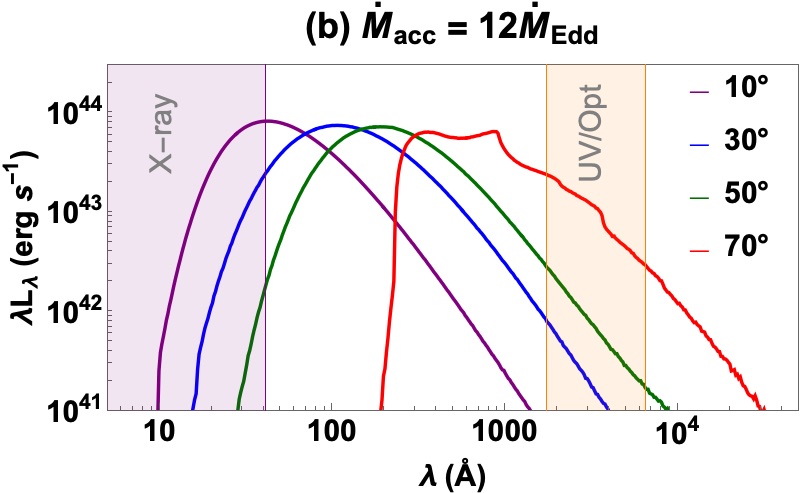}
        \figurenum{2b}
        \label{fig:spectra:acc12}
    \vspace{0.2cm}
   
     \includegraphics[width=0.44\textwidth]{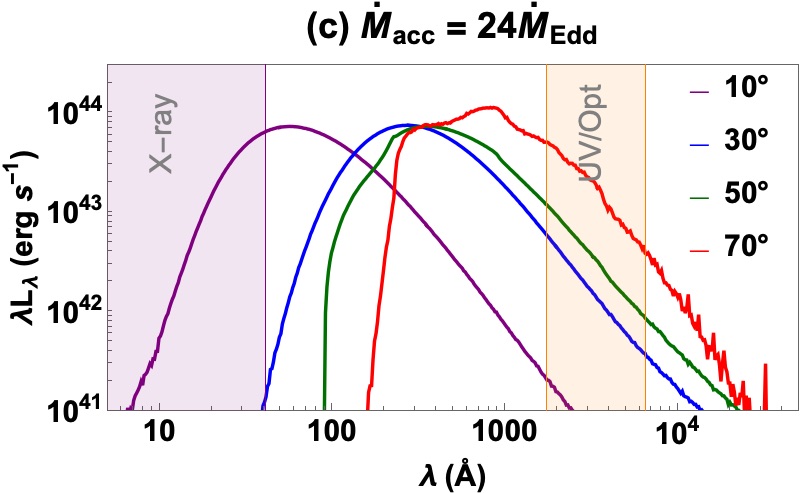}
    \figurenum{2c}
    \label{fig:spectra:acc24}
     \hspace{0.1cm}
    \includegraphics[width=0.44\textwidth]{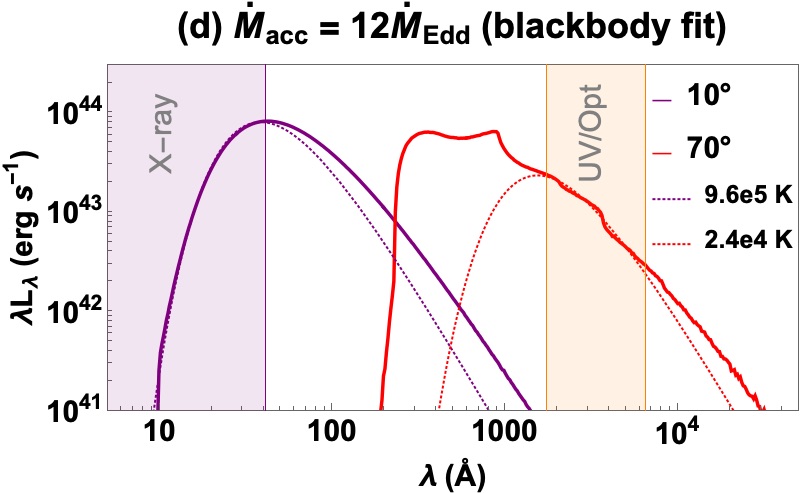}
    \figurenum{2d}
    \label{fig:spectra:fit}
    \vspace{0.2cm}
    
    \includegraphics[width=0.44\textwidth]{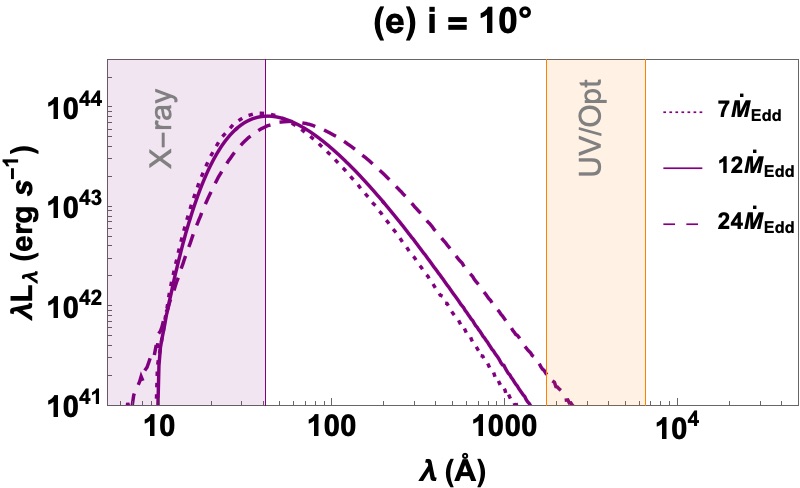}
    \figurenum{2e}
    \label{fig:spectra:i10}
     \hspace{0.1cm}
    \includegraphics[width=0.44\textwidth]{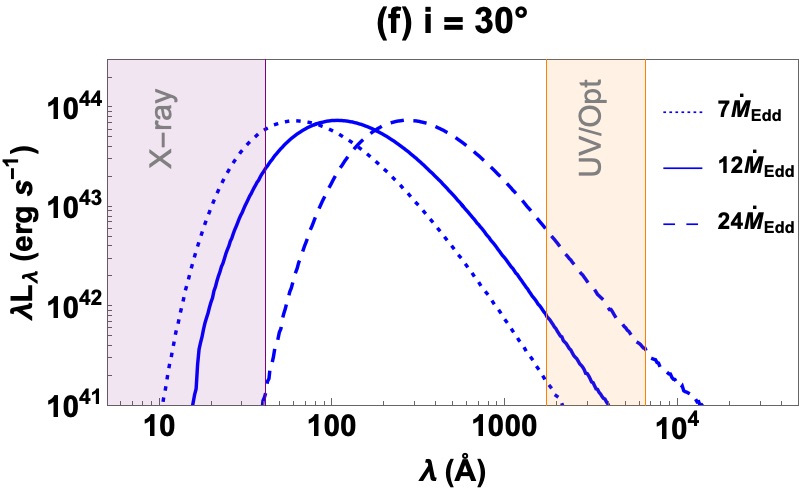}
    \figurenum{2f}
    \label{fig:spectra:i30}
     \vspace{0.2cm}

    \includegraphics[width=0.44\textwidth]{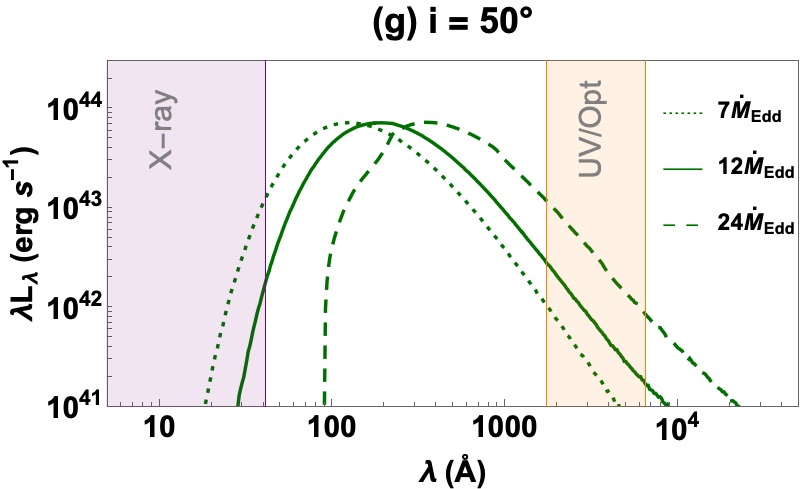}
    \figurenum{2g}
    \label{fig:spectra:i50}
    \hspace{0.1cm}
    \includegraphics[width=0.44\textwidth]{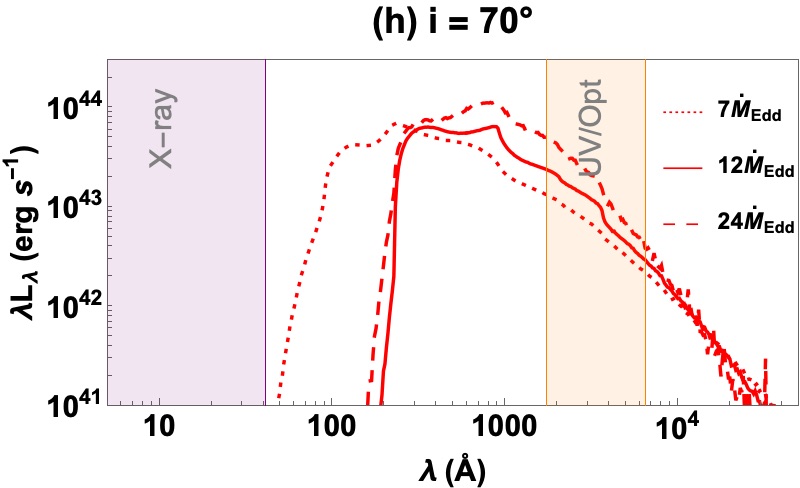}
    \figurenum{2h}
    \label{fig:spectra:i70}
    \hspace{0.1cm}

    \figurenum{2}
    \caption{\textbf{The simulated escaping spectra of the accretion disk at different accretion rates ($\dot{M}_{\rm acc}=$ 7, 12, and 24 $\dot{M}_{\rm Edd}$) and inclinations ($i=10^\circ, 30^\circ, 50^\circ, 70^\circ$).} The bolometric luminosity of the spectra $L_{\rm bol} = L_{\rm Edd}$ for all spectra. The purple shaded region indicates the X-ray band with an energy above 0.3 keV. The orange shaded region corresponds to \textit{Swift} UVOT band at 1700-6500 \AA. Panel (a)-(c) show the spectral evolution with inclination angles at fixed accretion rate. All spectra change from X-ray strong to UV/optical strong as the inclination goes from the polar direction to the disk direction. Panel (d) is the same as panel (b), but only includes the spectrum at $i=10^\circ$ with a blackbody spectrum fitting its X-ray continuum component, and the spectrum at $70^\circ$ with another blackbody spectrum fitting its UV/optical continuum component. Panel (e)-(h) show the spectral evolution with accretion rates at fixed inclinations. Three types of evolution can happen as accretion rate decreases: X-ray strong all the time (small inclination),  optical/UV strong at early time and X-ray rebrightening at late time (intermediate inclination), optical/UV strong all the time (large inclination). }
    \label{fig:spectra}
\end{figure}

We re-examine the emission viewing-angle dependence in the three new simulated disks. Fig. \ref{fig:spectra:acc7}, \ref{fig:spectra:acc12} and  \ref{fig:spectra:acc24} show how the escaped spectrum varies with inclination, when $\dot{M}_{\rm acc} = 7, 12, {\rm and} \ 24 \ \dot{M}_{\rm Edd}$ respectively. At any of these accretion rates, the SED still evolves from X-ray   to optically dominated as the inclination increases. However, the exact angle at which this transition occurs  depends on the accretion rate.
One can clearly observe that at low accretion rates $\approx {\rm few} \times \dot{M}_{\rm Edd}$, the escaped emission is dominated by X-ray emission at most inclinations unless the inclination angle is substantial (e.g., $i \gtrsim 70^{\circ}$ for  $\dot{M}_{\rm acc}= 7 \ \dot{M}_{\rm Edd}$), while at very high accretion rates $\approx {\rm few} \times 10 \ \dot{M}_{\rm Edd}$, the emanating radiation is mainly optically dominated except near the polar region (e.g., $i \lesssim 10^{\circ}$ for  $\dot{M}_{\rm acc}= 24 \ \dot{M}_{\rm Edd}$). 

We expect a TDE to be observed along a fix viewing angle during its entire evolution, unless the disk experiences some axial precession. Therefore, we  show the spectral evolution with $\dot{M}_{\rm acc}$ at fixed inclination angles in Fig. \ref{fig:spectra:i10}--\ref{fig:spectra:i70}. As the accretion rate declines after peak, the amount of reprocessing material is reduced, so the SED universally shifts towards lower wavelengths. However, the exact  behavior of the spectrum depends sensitively on the inclination angle. At very small inclinations ($i=10^{\circ}$), the TDE stays X-ray dominated throughout its evolution. At intermediate inclinations ($i=30^\circ$ and $50^{\circ}$), the TDE can be optically strong at early times (if $\dot{M}_{\rm acc}$ can be sufficiently large), and then rebrightens at X-ray energies at late time when the accretion rate diminishes. At very large inclinations ($i= 70^{\circ}$), the TDE stays UV/optically strong throughout its entire evolution.

While we have assumed that the bolometric luminosities of the escaped radiation always close to $L_{\rm Edd}$ in the analysis above, GRRMHD simulations show that the true escaped luminosity from super-Eddington disks can exceed the Eddington limit, with more flux leaking out through the funnel \citep[e.g.,][]{Sadowski16}. Therefore, we also calculated the spectra when the escaped radiation has a bolometric luminosity $L_{\rm bol}= 10\% \ \dot{M}_{\rm acc} c^2$, which are shown in Fig. \ref{fig:spectrahighL}. 
When the reprocessing is driven by adiabatic expansion, the SED shape stays unchanged while the magnitude of the flux scales with the luminosity.
When the reprocessing mechanism is driven by absorption and Comptonization, increasing the luminosity makes the gas more ionized and reduces bound-free and free-free absorption, which shifts the spectral energy peak to slightly higher energies. However, the spectral shape is rather insensitive to the luminosity. 
In general, the setting of the luminosity within the explored range does not alter how the escaped radiation depends on the viewing angle and accretion rate.

\subsection{Comparison with observations: blackbody luminosity, effective temperature, and photosphere radius} \label{sec:observables}

\begin{figure}[h!]
    \centering
    \includegraphics[height=0.25\textheight]{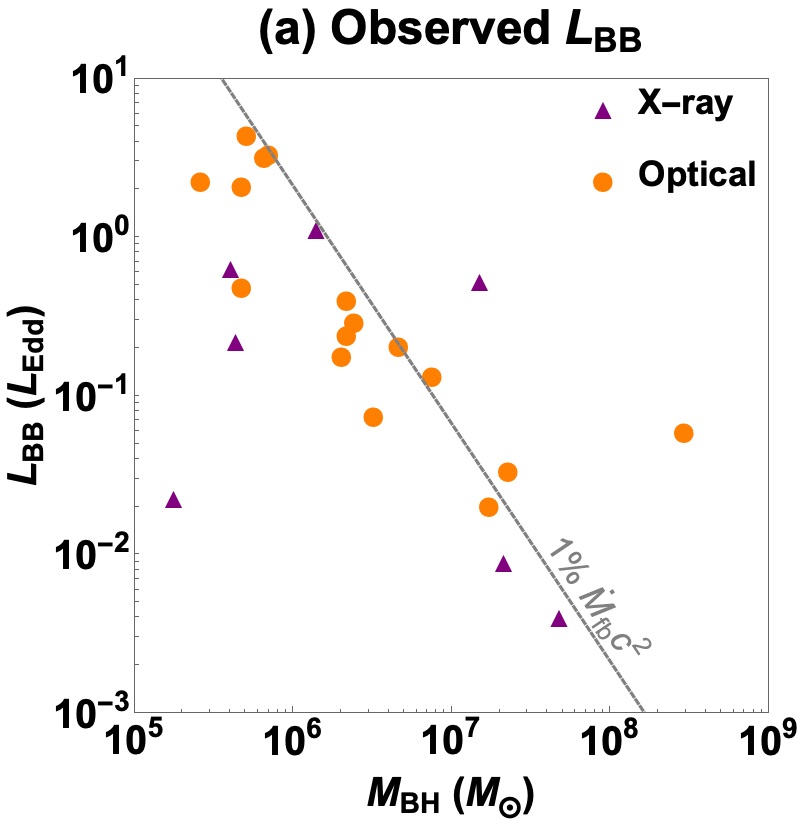}
    \figurenum{3a}
    \label{fig:obs:L}
    \hspace{0.1cm}
    \includegraphics[height=0.25\textheight]{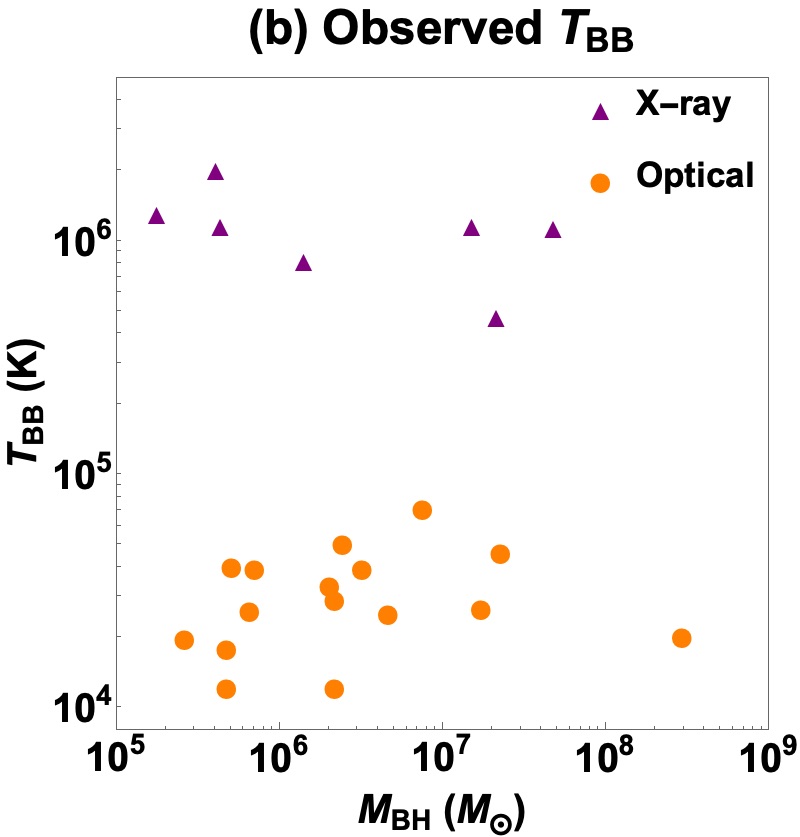}
    \figurenum{3b}
    \label{fig:obs:T}
    \hspace{0.1cm}
    \includegraphics[height=0.25\textheight]{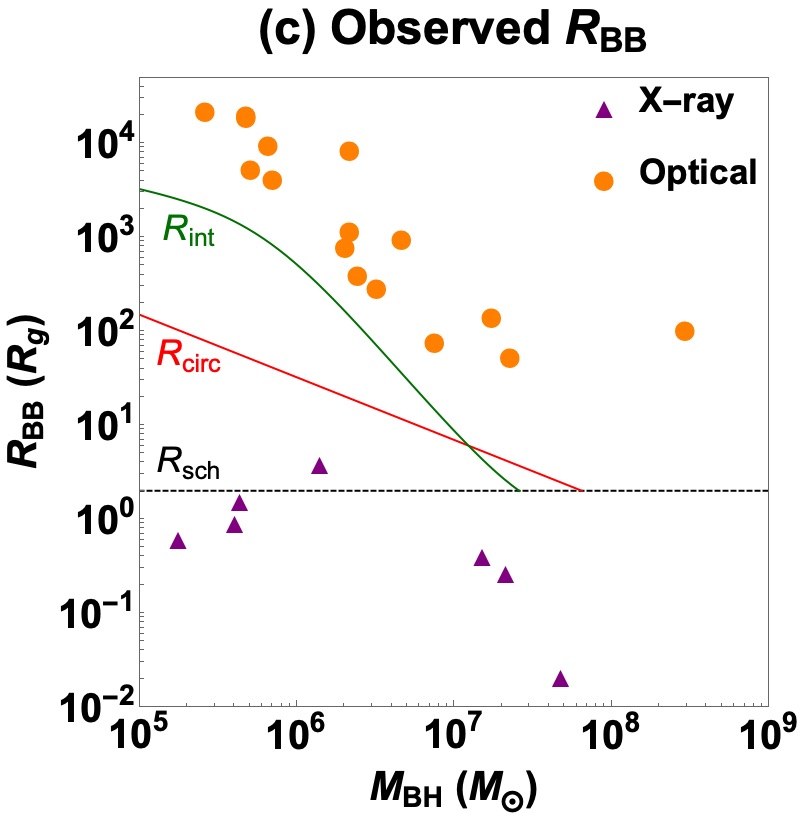}
    \figurenum{3c}
    \label{fig:obs:R}
    \vspace{0.3cm}
    
    \includegraphics[height=0.25\textheight]{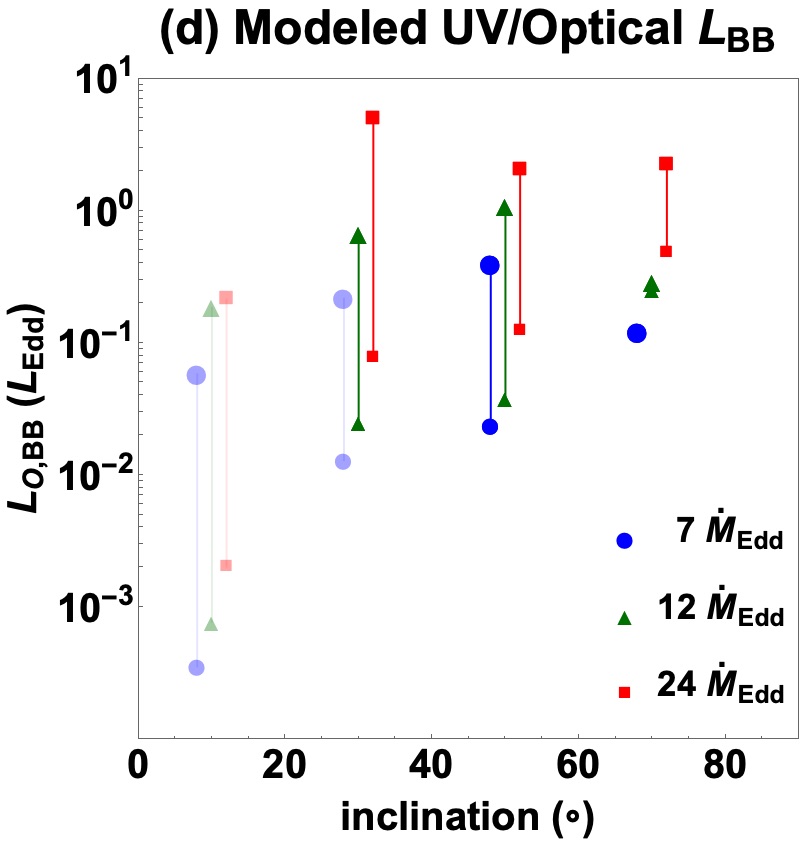}
    \figurenum{3d}
    \label{fig:model:Lo}
    \hspace{0.1cm}
    \includegraphics[height=0.25\textheight]{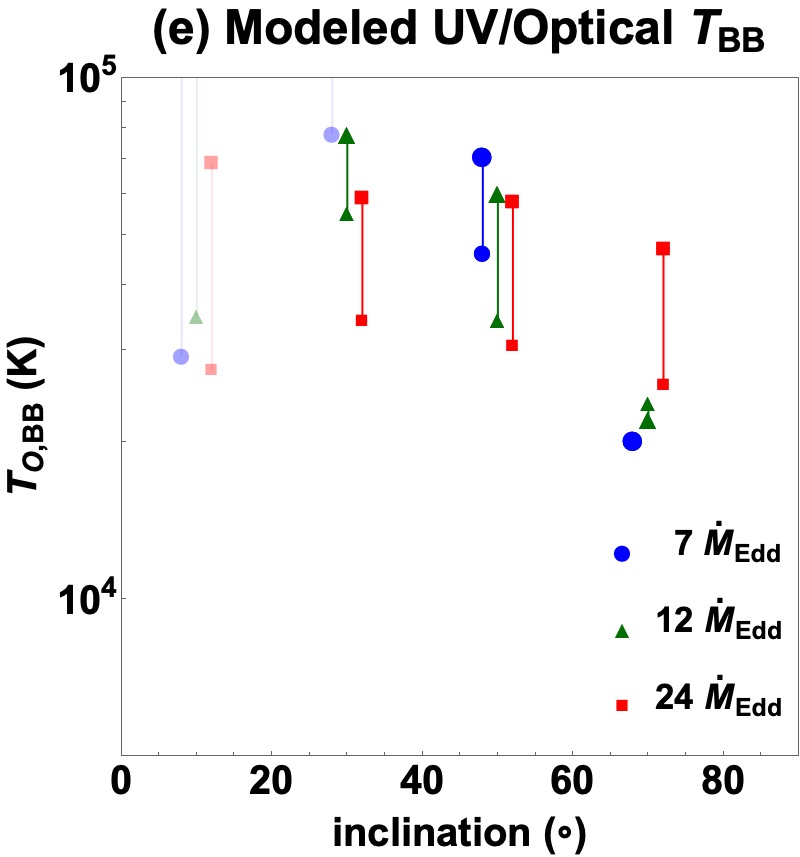}
    \figurenum{3e}
    \label{fig:model:To}
    \hspace{0.1cm}
    \includegraphics[height=0.25\textheight]{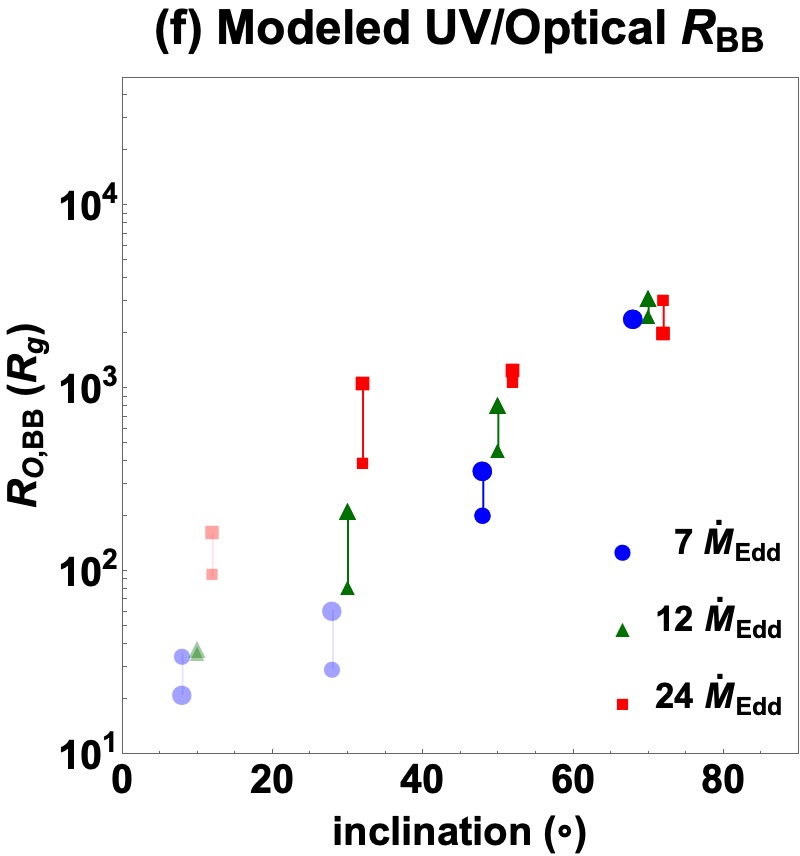}
    \figurenum{3f}
    \label{fig:model:Ro}
    \vspace{0.3cm}
     
    \includegraphics[height=0.25\textheight]{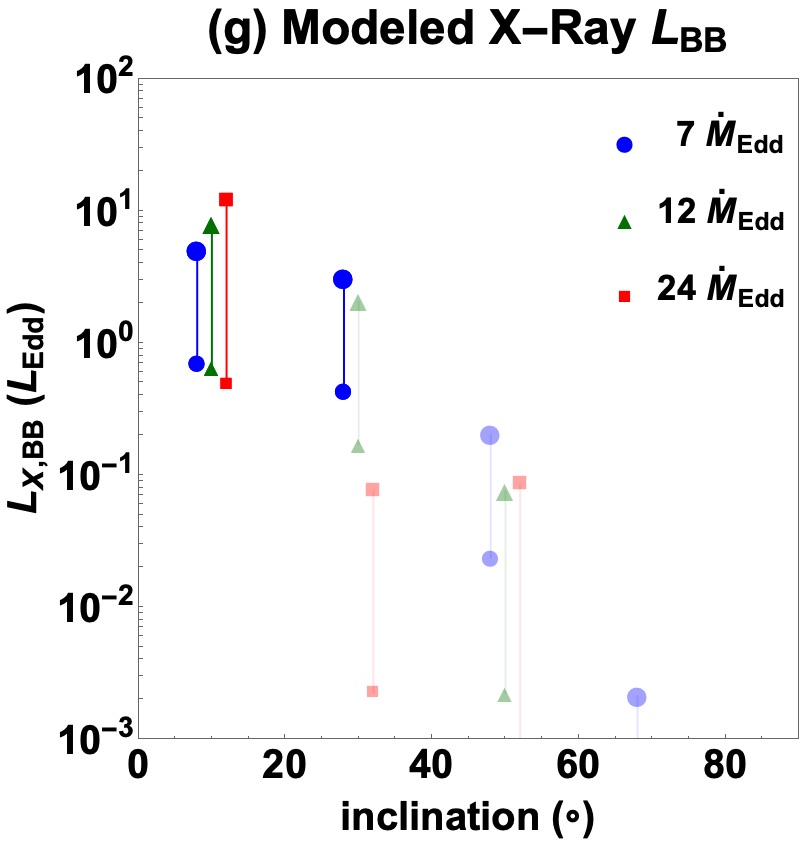}
    \figurenum{3g}
    \label{fig:model:Lx}
    \hspace{0.1cm}
    \includegraphics[height=0.25\textheight]{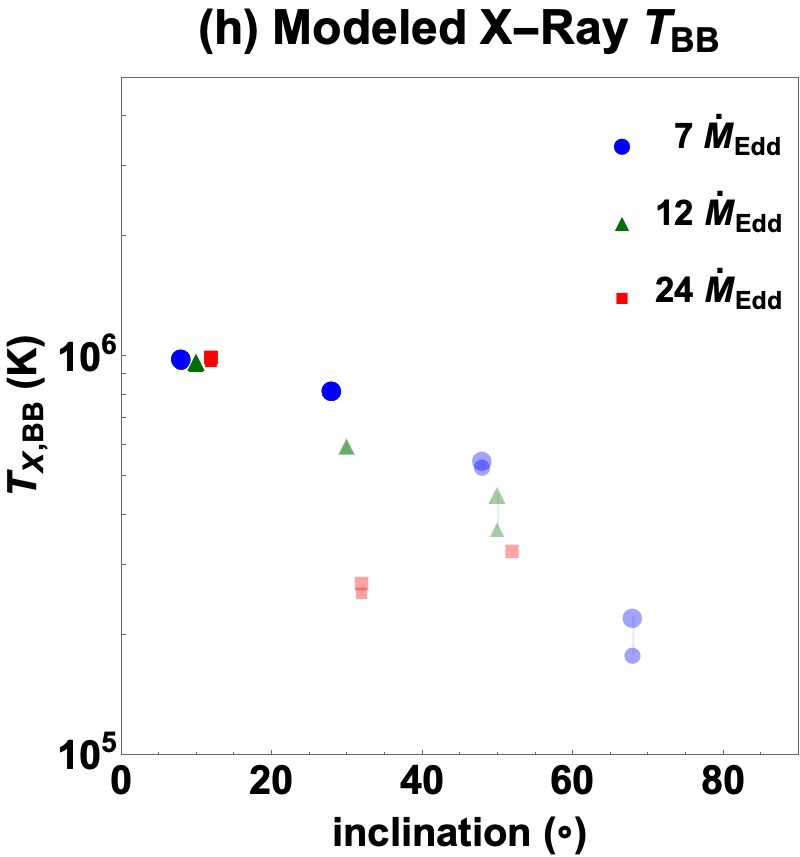}
    \figurenum{3h}
    \label{fig:model:Tx}
    \hspace{0.1cm}
    \includegraphics[height=0.25\textheight]{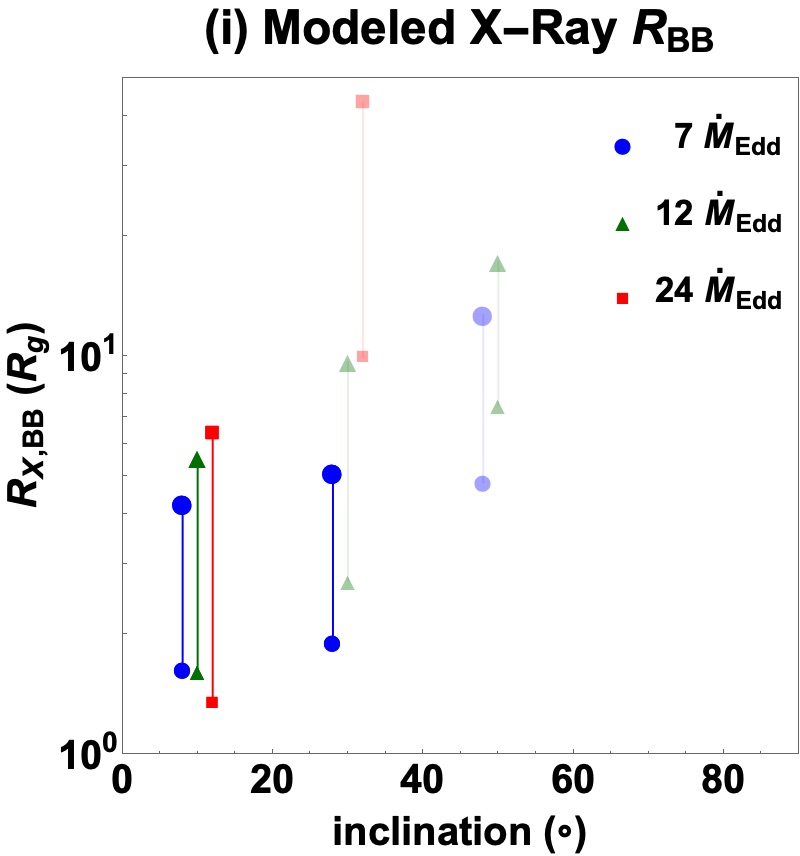}
    \figurenum{3i}
    \label{fig:model:Rx}
    \figurenum{3}
    \caption{\textbf{Comparison between the observed and modeled TDE blackbody luminosity, temperature and radius.} Upper row (a)-(c): The observed quantities vs. $M_{\rm BH}$ for 16 optically-selected TDEs (orange dots) and 7 X-ray-selected TDEs (purple triangles). (a) The observed $L_{\rm BB}/L_{\rm Edd}$ has a clear trend with $M_{\rm BH}$. A gray line showing $L=1 \% \ \dot{M}_{\rm fb} c^2$, with $\dot{M}_{\rm fb}$ being the fallback rate of a $0.1 M_\odot$ star is overplotted to guide the eye. (b) X-ray TDEs have temperatures of $10^5-10^6$K, and optical TDEs typically have lower temperature at few$\times10^4$K. (c) Optical TDEs have $R_{\rm BB}$ larger than the circularization radius (red curve) or the stream self-intersection radius (green curve) (both calculated using a $0.1 M_\odot$ star). X-ray TDEs can sometimes have $R_{\rm BB}$ smaller than the black hole Schwarzschild radius (black line).
    Middle row (d)-(f) and lower row (g)-(i): The inferred quantities based on the blackbody radiation spectrum fitting the simulated spectra in the UV/Optical band or X-ray band respectively vs. inclination angle $i$. Different symbols are used to mark different accretion rates: $7\dot{M}_{\rm Edd}$ (blue circle), $12\dot{M}_{\rm Edd}$ (green triangle) and $24 \dot{M}_{\rm Edd}$ (red square). Vertical lines connect the values calculated with an escaped luminosity of $L_{\rm bol}=L_{\rm Edd}$ (smaller symbol size) and $L_{\rm bol} =10\%\dot{M}_{\rm acc} c^2$ (larger symbol size) to indicate possible ranges. In (d)-(f), TDEs with $L_{\rm O, BB} < L_{\rm X, 0.3-10 keV}$ are masked with low opacity to indicate they are less likely to be selected optically. Similarly, in (g)-(i), TDEs with $L_{\rm O, BB} > L_{\rm X, 0.3-10 keV}$ are masked with low opacity to indicate they are less likely to be selected by X-ray instruments. The blackbody luminosity, temperature and radius inferred from our modeling, to the first order, reproduce the observed ones. 
 }
    \label{fig:observable}
\end{figure}

\noindent In this section, we compare our model predictions to the observed properties of TDEs. We start from the TDE catalog in \citet{Gezari21review} which also lists the observed blackbody luminosities and temperatures, and then only include TDEs which have their masses estimated from the $M-\sigma$ relation as in \citet{Wong22}. This gives us 7 X-ray-selected TDEs (Table \ref{tab:Optical_TDE}) and 16 optically-selected TDEs (Table \ref{tab:XrayTDE}). We plot their observed blackbody luminosities, effective temperatures and photosphere radii as functions of $M_{\rm BH}$ in Fig. \ref{fig:obs:L} - \ref{fig:obs:R}. The observed $L_{\rm BB}$ varies between $10^{-3}-10 \ L_{\rm Edd}$, and can exceed $L_{\rm Edd}$ when $M_{\rm BH} \lesssim 10^6 M_\odot$. Interestingly, $L_{\rm BB}$ commonly has a dependence on $M_{\rm BH}$ following the fallback rate trend: $\dot{M}_{\rm fb}/\dot{M}_{\rm Edd}\propto M^{-3/2}_{\rm BH}$. The observed $T_{\rm BB}$ clearly depends on whether the TDE is optically-selected ($\approx {\rm few}\times10^4$ K) or X-ray-selected ($\approx 10^5-10^6$ K). The observed $R_{\rm BB}$, calculated from $L_{\rm BB}$ and $T_{\rm BB}$ using the Stefan-Boltzmann law, also has a bimodal distribution. The optically-selected TDEs have $R_{\rm BB}$ reaching $10^2-10^4 R_g$ and always exceeding both the circularization radius $R_{\rm circ} = 2 R_t$ and the stream self-intersection radius $R_{\rm int}$ \citep{Dai15}. The X-ray-selected TDEs, however, sometimes have $R_{\rm BB}$ even within the event horizon, which might be caused by absorption in circumnuclear medium.

As the observed properties of TDEs are usually inferred from their spectra in monochromatic bands, we also fit the simulated spectra in either the X-ray or UV/optical band with blackbody  spectra. For example, in Fig. \ref{fig:spectra:fit}, we show the simulated escaped spectra for $\dot{M}_{\rm acc}= 12 \dot{M}_{\rm Edd}$ at $i=10^\circ$ and $70^\circ$ and the blackbody radiation fits to the spectra in the X-ray band (0.3-10 keV) or the part of the spectrum in the UV/optical band (the 170-650 nm {\it Swift} UVOT band). It can be seen that the X-ray and UV/optical emission can be individually well approximated by blackbody radiation.
Similarly, for each spectrum (at $\dot{M}_{\rm acc}=$ 7, 12, \rm or 24 $\dot{M}_{\rm Edd}$ and inclination $i=10^\circ, 30^\circ, 50^\circ, \rm or \ 70^\circ$), we obtain two blackbody radiation fits in the X-ray band and the UV/optical band.  The escaped bolometric luminosity is  assumed to be varying between $L_{\rm Edd}$ and $0.1\% \ \dot{M}_{\rm acc} c^2$. The luminosities, temperatures and radii of the best-fit blackbody radiation spectra are are listed in Table \ref{tab:modelobservable} and plotted in Fig. \ref{fig:model:Lo} - \ref{fig:model:Ro} (UV/optical fit) and Fig. \ref{fig:model:Lx} - \ref{fig:model:Rx} (X-ray fit).
We further categorize whether a modeled TDE spectrum is a X-ray strong or optically strong by comparing the luminosity in the X-ray band (0.3-10 keV) and the blackbody luminosity inferred from the UV/optical band. We compare the model predictions to the observed properties as below:

\hangindent=0.5cm
\hangafter=0
1. \textit{Luminosity}: The modeled $L_{\rm BB}$ mostly lies between 0.01-few $L_{\rm Edd}$. The simulated spectrum usually peaks in EUV and has a broader shape than a single-temperature blackbody spectrum. Therefore, the inferred blackbody luminosity ($L_{\rm O, BB}$ or $L_{\rm X, BB}$) is always smaller than the bolometric luminosity $L_{\rm bol}$ of the escaped radiation. 
$L_{\rm O, BB}/L_{\rm bol}$ for optically-strong TDEs and $L_{\rm X, BB}/L_{\rm bol}$ for X-ray-strong TDEs is typically few$\times(1-10)\%$.  This naturally explains the missing energy problem in TDE \citep{Stone16, Lu18, Mockler21} -- the majority of the energy is emitted in the EUV band which is difficult to be detected.

\hangindent=0.5cm
\hangafter=0
2. \textit{Temperature}: Our modeling reproduces the bimodal distribution of TDE temperatures, i.e., optically-strong TDEs have temperature $T_{\rm O, BB} \approx {\rm few}\times10^4 K$, and X-ray-strong TDEs have temperature $T_{\rm X, BB} \lesssim 10^6 K$. Furthermore, our modeling shows that the inferred temperatures of optical TDEs are not highly sensitive to accretion rates or observer inclination, which explains why TDEs have relatively constant $T_{\rm BB}$ throughout the evolution \citep{vanVelzen20review}.

\hangindent=0.5cm
\hangafter=0
3. \textit{Photosphere radius}: The optically-strong TDEs have far larger photosphere radii ($R_{\rm O, BB}\approx 10^2-10^4 R_g$) than X-ray-strong TDEs ($R_{\rm X, BB}\approx  {\rm few}\times R_g$). A comparison between the observed $R_{\rm O, BB} = 10^3-10^4 R_g$ and the modeled $R_{\rm O, BB}$ suggests that optically-selected TDEs are either commonly observed from large inclinations, or have $L_{\rm bol}> L_{\rm Edd}$. The distinction between the observed and modeled $R_{\rm X, BB}$ is possibly accounted by the absorption of X-rays in the circumnuclear medium.\\

We note that our predictions for the X-ray quantities, in particularly in the $i=10^\circ$ bin, are sensitive to the setting of the radiative transfer calculations. Here we always inject a blackbody spectrum with a constant $T=10^6K$, so our predicted $T_{\rm X, BB}$ at small inclinations also fall into a very narrow range. However, the temperatures at the center of accretion disks generally increase with increasing $\dot{M}_{\rm acc}$ and decreasing $M_{\rm BH}$, which will induce more variance to the observed X-ray temperatures.  Also, for the setting with $L=L_{\rm Edd}$, the predicted $L_{\rm X, BB}$ decreases as $\dot{M}_{\rm acc}$ increases, as a result of a constant bolometric luminosity and more reprocessing from X-ray to UV/optical emissions at higher densities. However, simulations show that the radiation fluxes leaking out through the funnel are not Eddington-limited \citep{McKinney15, Dai18}. Therefore, we expect that X-ray luminosities should scale positively with accretion rates, as illustrated by the $L=0.1 \ \dot{M}_{\rm acc} c^2$ case.

\subsection{Temporal evolution of TDEs}\label{sec:evolution}

\begin{figure}[h!]
    \centering
    \includegraphics[height=0.33\textheight]{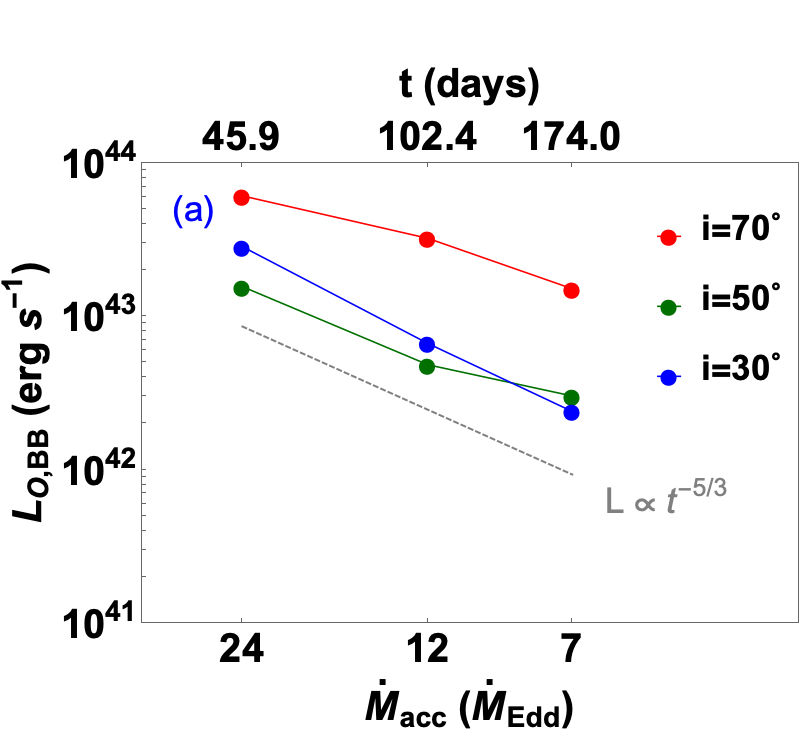}
    \figurenum{4a}
    \label{fig:evo:L}
    \hspace{0.2cm}
    \includegraphics[height=0.33\textheight]{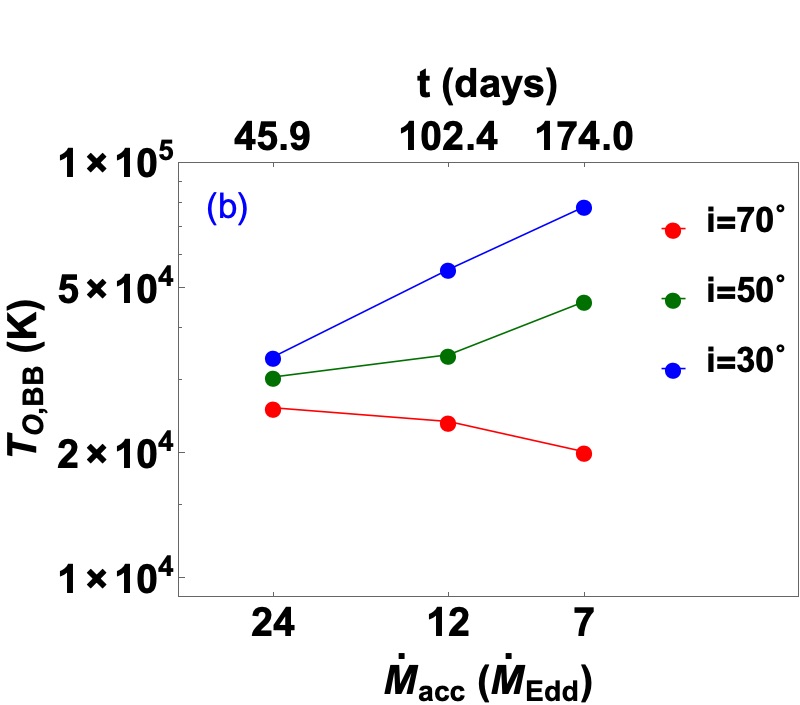}
    \figurenum{4b}
    \label{fig:evo:T}
    \vspace{0.3cm}
    
    \includegraphics[height=0.33\textheight]{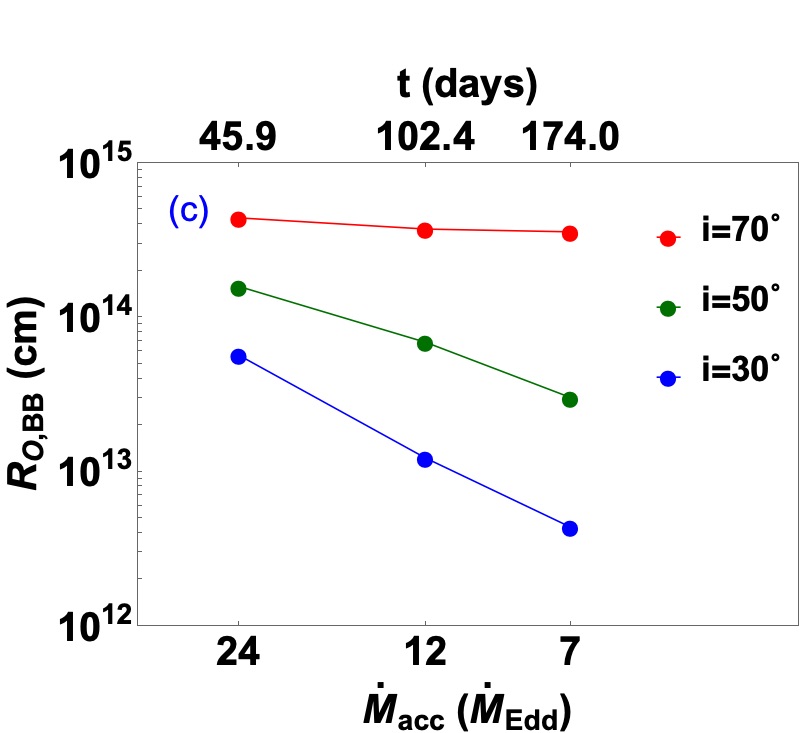}
    \figurenum{4c}
    \label{fig:evo:L}
    \hspace{0.3cm}
    \includegraphics[height=0.33\textheight]{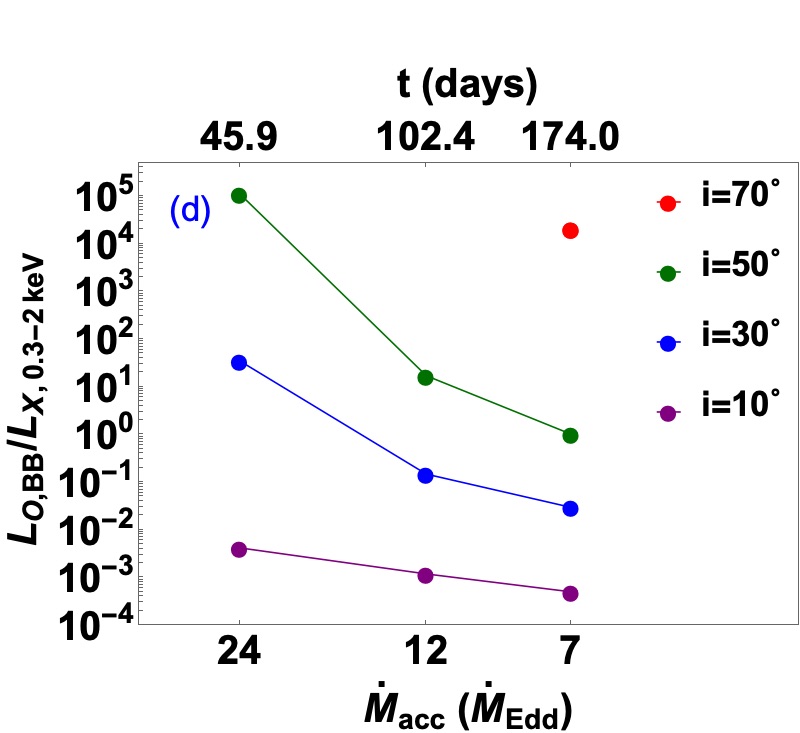}
    \figurenum{4d}
    \label{fig:evo:LoLx}
     \figurenum{4}
    \caption{\textbf{The post-peak temporal evolution of the modeled TDE UV/optical  luminosity (a), temperature (b), radius (c), and the ratio between the UV/optical and X-ray luminosity (d).} Different colors denote different inclination angles. The escaped radiation has luminosity $L_{\rm bol} = L_{\rm Edd}$ for all curves. The lower axis shows the accretion rate and the upper axis shows the corresponding time elapsed since the peak, assuming a solar-type star disrupted by a $10^6 M_\odot$ black hole. In (a)-(c), we do not include the evolution $i=10^\circ$, where the event is always X-ray strong. In (a) the gray line shows the trend of $t^{-5/3}$ to guide the eye. In (d) the X-ray luminosity includes only the flux in the 0.3-2 keV band for direct comparison with observations. Also at $i=70^\circ$ the X-ray luminosities at the two higher accretion rates are negligible.
 }
    \label{fig:evolution}
\end{figure}

\noindent We show in Fig. \ref{fig:evolution} the evolution of the modeled  luminosity, temperature, radius, as well as the ratio of optical to X-ray luminosity, as functions of the accretion rate. In order to connect the snapshots at an specific accretion rate to the temporal evolution of TDEs, we assume $\dot{M}_{\rm fb} (t) = \dot{M}_{\rm acc} (t) +  \dot{M}_{\rm wind}(t)$. This assumption is valid only if the fallback timescale dominates over other timescales, such as the disk viscous timescale and the photon diffusion/advection timescales. The exact conversion from $\dot{M}_{\rm fb}$ to $t$ depends on the mass of the MBH, the properties of the disrupted star, and the impact parameter \citep{Law-Smith20}. Focusing on the post-peak evolution, the three disk simulations correspond to 45.9, 102.4 and 174 days after the peak of the flare, assuming a solar-type star is fully disrupted by a $10^6M_\odot$ black hole. Calculations of various timescales are given in Appendix \ref{app:timescale}.

As the accretion rate drops from $24\dot{M}_{\rm Edd}$ to $7\dot{M}_{\rm Edd}$, the optical luminosity also decreases, and the UV/optical light curve roughly follows the canonical $t^{-5/3}$ decay. As discussed in the previous section, the fitted blackbody temperature stays rather constant. Interestingly, at large inclinations $T_{\rm O, BB}$ slightly decreases with declining $\dot{M}_{\rm acc}$, while at small to intermediate inclinations $T_{\rm O, BB}$ shows the opposite trend. This can provide an explanation to the different observed evolution of $T_{\rm O, BB}$ \citep{vanVelzen20review}. As a result, the photosphere radius shrinks as $\dot{M}_{\rm acc}$ decreases, with a faster evolution at smaller inclinations.

The ratio between UV/optical luminosity ($L_{\rm O, BB}$) and X-ray luminosity ($L_{\rm X, 0.3-2 keV}$) also decreases as the accretion level drops and the amount of obscuring material is reduced. 
The fastest $L_{\rm O, BB}/ L_{\rm X, 0.3-2 keV}$ evolution is observed at intermediate inclinations, and especially if the TDE has a high accretion rate at peak. In such cases, we expect to see strong X-ray re-brightening of initially optically-strong TDEs, as reported in \citet{Gezari17}. The X-ray luminosity reaches the same level as the UV/optical luminosity at $t\gtrsim 100$ days after peak, and possibly even later if the disk formation or viscous timescale is long. At very small inclinations, the event is always X-ray strong. At very large inclinations, the event should stay optically strong for a long period, although it is theoretically predicted that the disk should eventually become geometrically thin and emit mostly in X-rays/UV when the accretion level drops to around Eddington \citep{Shakura73}.

\section{Summary and future work}
\label{sec:summary}

\noindent
Inspired  by the unified model for TDEs proposed by in \citet{Dai18}, we carry out three addtional  3D GRRMHD simulations of TDE super-Eddington accretion flow at different Eddington ratios, and conduct radiative transfer calculations to obtain the emanating spectra. Based on the results, we further propose a dynamical unified model which can explain the diversity and evolution of TDEs:

\begin{itemize}
\item The viewing angle of the observer with respect to the orientation of the disk is the most important parameter in determining whether we observe either an X-ray or an optical bright TDE. At small inclinations, X-rays can escape from the funnel of the super-Eddington disk. At large inclinations, X-rays are mostly reprocessed into UV/optical radiation by the geometrically and optically thick wind and disk.

\item The blackbody radiation fits of the TDE super-Eddington disk spectra produce effective temperature, blackbody luminosity and photosphere radius distributions consistent with the observed ones. Most radiative energy escapes in the EUV range, and only a few to a few tens of percentage of radiative energy can be detected, which provides a solution to the TDE missing energy problem.

\item  The observed diversity of the emission from different TDEs can be associated with the different Eddington ratios of their accretion rates, $\dot{M}_{\rm acc} = {\rm few} \times (1-10) \ \dot{M}_{\rm Edd} $, at the flare peak conditions. In general, higher accretion levels induce larger (fitted blackbody) luminosities and larger photosphere radii, but do not significantly change the fitted effective temperatures.

\item The early-time evolution ($t\approx 100$ days after peak) of optical TDEs can be explained by this reprocessing model. As the luminosity drops by about 0.5-1 order of magnitude, the fitted temperature slightly increases at small to intermediate inclinations or decreases at large inclinations.

\item The evolution of the optical-to-X-ray flux ratio also depends sensitively  on the viewing angle. At large inclinations, the TDE stays UV/optically strong for a very long time. At intermediate inclinations, we expect to see the fastest X-ray rebrightening, and $L_O/L_X$ reaches unity at few hundred days. At small inclinations, the TDE is always X-ray strong. The exact evolution timescale also depends on the accretion rate at peak, which further depends on the black hole and stellar parameters.
\end{itemize}

\vspace{0.1cm}
In this study we focus on understanding the disk continuum emission. We will investigate the disk spectroscopic features in a future work. Also, while our GRRMHD simulations are conducted in 3D, we only do post-processing along different inclination bins by assuming the profiles are spherically symmetric, which means that photons have an overall radial motion. Implementing 2D or 3D radiative transfer calculations can allow us to study the reprocessing of photons along more realistic paths. Furthermore, the evolution of the disk and jet emissions can be studied in more details by conducting simulations covering a more complete parameter space for black hole accretion rate and spin. 

The unified model we have proposed for TDEs also sheds light on other types of black hole accretion systems, especially other super-Eddington sources such as ultra-luminous X-ray sources, narrow line Seyfert 1 galaxies, changing-look AGN and quasars in the early universe. Works like this illustrate that novel simulations are crucial for making direct comparisons between the observed emissions and model predictions, which offers important insights to how black holes grow through cosmic time, use accretion as a source of energy to power outflows and radiation, and give feedback to their surrounding environment.

\acknowledgments We thank K. Auchettl, M. Bulla, S. Gezari, D. Kasen, G. Leloudas, B. Mockler, and N. Stone for useful discussions. LT, TK and LD acknowledge the support from the Hong Kong Research Grants Council (HKU27305119, HKU17304821) and the National Natural Science Foundation of China (HKU12122309). ER-R is grateful for support from the Heising-Simons Foundation, NSF (AST-1615881, AST-1911206 and AST-1852393), Swift (80NSSC21K1409, 80NSSC19K1391) and Chandra (GO9-20122X). The simulations carried out for this project were performed on the HPC computing facilities offered by ITS at HKU and the Tianhe-2 supercluster.

\bibliographystyle{yahapj}
\bibliography{references}

\appendix
\counterwithin{figure}{section}
\counterwithin{table}{section}

\section{Properties of simulated disks}
\label{app:simulation}
\noindent A few more time-averaged quantities of the three simulated disks are listed in Table \ref{tab:morequantities}. $\Phi_{H}$ is the normalised magnetic flux at horizon $\Phi_{H}$ \citep{Tchekhovskoy11}, with $\Phi_{H}\gtrsim 30-40$ being the condition for MAD disks.
The net accretion efficiency $\eta_H$ evaluates how much rest-mass energy going into the black hole is converted to out-going energy near the event horizon. $L_{\rm{} jet}$ is the total power of the relativistic jet at $r = 5500 r_g$, most of which is in the form of electromagnetic energy. $\Gamma_{\rm{} jet, max}$ is the maximum Lorentz factor of the jet. $L_K$ is the thermal+kinetic luminosity of the wind calculated at $r \sim 5500 r_g$. 
$\alpha_{\rm{} eff}$ is the effective $\alpha$-parameter of the disk as defined in \citet{McKinney12}. $R_{\rm eq}$ indicates the radius within which the disk inflow equilibrium has been established. In our simulations, the winds are launched mostly from the inner disk regions and have traveled beyond the photospheres at most inclination angles.

\begin{deluxetable*}{cccccccccccc}[h]
\tablecaption{More disk quantities \label{tab:morequantities}}
\tablewidth{0pt}
\tablehead{
\colhead{Model}  & \colhead{$\dot{M}_{\rm acc}\ (\dot{M}_{\rm{}Edd})$}     & \colhead{$\Phi_{H}$}  & \colhead{$\eta_{H}\ (\%)$} & \colhead{$L_{\rm{} jet}\ (L_{\rm{}Edd})$} &  \colhead{$\Gamma_{\rm jet, max}$}   &  $L_K\ (L_{\rm Edd})$ & $\alpha_{\rm eff}$ & $R_{\rm eq} \ (R_g) $
}
\startdata
M6a08-1	&   7   &  34.5	&	35.5    &   1.1   &  2.27  &    0.5   &     2.7 & 340    \\
M6a08-2	&    12    &  39	&	40.1    &   2.3   &  2.39  &    1.0   &     3.3 & 320 \\
M6a08-3	&   24   &  55	&	69.3    &   9.7   &  2.66  &    4.8  &   2.3  &   230 \\
\enddata
\end{deluxetable*}

\section{Simulated spectra at higher luminosity}
\label{app:spectraLlarge}

\begin{figure}[h!]
    \centering
    \includegraphics[width=0.44\textwidth]{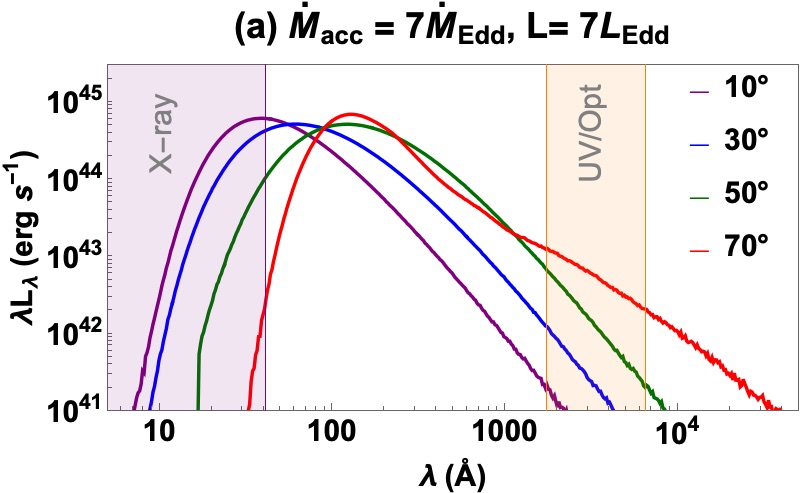}
    \label{fig:spectra:acc7highL}
    \hspace{0.1cm}
    \includegraphics[width=0.44\textwidth]{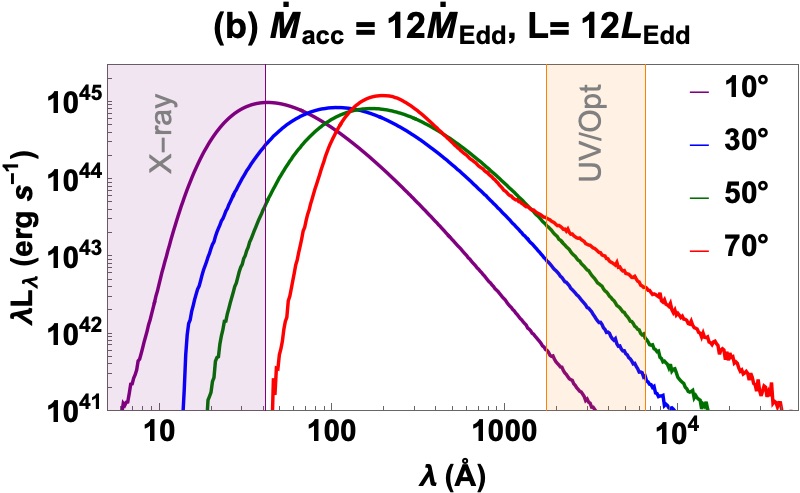}
    \label{fig:spectra:acc12highL}
    \vspace{0.2cm}
    
    \includegraphics[width=0.44\textwidth]{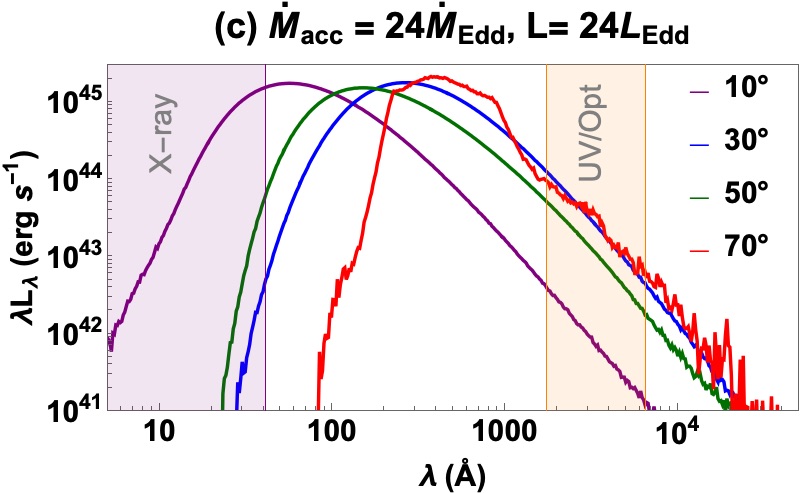}
    \label{fig:spectra:acc24highL}
    
    \caption{The simulated escaping spectra of the accretion disk at different accretion rates and inclinations, similar to Fig. \ref{fig:spectra:acc7}-\ref{fig:spectra:acc24}, except that the bolometric luminosity of the spectra $L_{\rm esc} = 10\% \times \dot{M}_{\rm acc} c^2$. 
 }
    \label{fig:spectrahighL}
\end{figure}

\section{List of observed TDEs reported in literature}
\label{app:TDElist}
\noindent For plotting Fig. \ref{fig:obs:L}-\ref{fig:obs:R} we use 16 optically-selected TDEs and 7 X-ray-selected TDEs. For completeness we list their names and relevant parameters as reported from pervious literature in Table \ref{tab:Optical_TDE} and \ref{tab:XrayTDE}. 

\begin{table}[h]
    \centering
    \caption{Names and parameters of optical TDEs}
    \begin{tabular}{lcccc}
    Name & $\log \ M_{\rm BH} \ (M_\odot) $  &   $\log \ L_{\rm BB} \ (\rm erg \ s^{-1}) $ &  $\log \ T_{\rm BB} \ (\rm K)$ &    log \ $R_{\rm BB}$ ($R_g$)\\
    \tableline
                ASASSN-14ae &           5.42 &    43.87 &  4.29 &    4.34 \\
           ASASSN-14li &           6.31 &    43.66 &  4.52 &    2.88 \\
           ASASSN-15lh &           8.47 &    45.34 &  4.30 &    2.00 \\
           ASASSN-15oi &           5.71 &    44.45 &  4.60 &    3.72 \\
 AT2018dyb/ASASSN-18pg &           6.67 &    44.08 &  4.40 &    2.97 \\
 AT2018hyz/ASASSN-18zj &           5.68 &    44.10 &  4.25 &    4.27 \\
             GALEXD1-9 &           6.51 &    43.48 &  4.59 &    2.45 \\
           GALEXD23H-1 &           6.39 &    43.95 &  4.70 &    2.59 \\
            GALEXD3-13 &           7.36 &    43.98 &  4.66 &    1.71 \\
             iPTF-15af &           6.88 &    44.10 &  4.85 &    1.87 \\
            iPTF-16axa &           6.34 &    43.82 &  4.46 &    3.05 \\
              PS1-10jh &           5.85 &    44.47 &  4.59 &    3.61 \\
             PTF-09axc &           5.68 &    43.46 &  4.08 &    4.29 \\
             PTF-09djl &           5.82 &    44.42 &  4.41 &    3.97 \\
              PTF-09ge &           6.34 &    44.04 &  4.08 &    3.92 \\
             SDSS-TDE1 &           7.24 &    43.64 &  4.42 &    2.14 \\ \tableline
    \end{tabular}
\label{tab:Optical_TDE}
\end{table}

\begin{table}[h]
    \centering
    \caption{Names and parameters of X-ray TDEs}
\begin{tabular}{lcccc}
    Name &  $\log \ M_{\rm BH} \ (M_\odot) $  &  $\log \ L_{\rm BB} \ (\rm erg \ s^{-1}) $ &  $\log \ T_{\rm BB} \ (\rm K)$ &   log \ $R_{\rm BB}$ ($R_g$)\\
    \tableline
 SDSSJ1323+48 &  6.15 &  44.30 &  5.91 &  0.80 \\
 SDSSJ1201+30 &  7.18 &  45.00 &  6.06 & -0.29 \\
   RXJ1624+75 &  7.68 &  43.38 &  6.05 & -1.57 \\
   RXJ1420+53 &  7.33 &  43.38 &  5.67 &  0.03 \\
      RBS1032 &  5.25 &  41.70 &  6.11 & -0.13 \\
 3XMMJ1521+07 &  5.61 &  43.51 &  6.30 & -0.02 \\
 3XMMJ1500+01 &  5.64 &  43.08 &  6.06 &  0.29 \\ \tableline
\end{tabular}
    \label{tab:XrayTDE}
\end{table}

\section{Modeled TDE Parameters}
\label{app:modelvalue}

\noindent The values of the modeled observable as in Fig. \ref{fig:model:Lo}-\ref{fig:model:Rx} are listed in Table \ref{tab:modelobservable}.

\begin{table}[h]
\renewcommand{\arraystretch}{1.8}
\centering
\caption{Modeled TDE observables}
\vspace{0.2cm}
\begin{tabular}{l p{0.8cm} ccccccccc}
 \tableline
 $\dot{M}_{\rm acc}$  &  $L_{\rm bol}$  &  inc   &  log $L_{\rm O, BB}$  &  $L_{\rm O, BB}/L_{\rm bol}$ &  log $T_{\rm O, BB}$ &  log $R_{\rm O, BB}$ &  log $L_{\rm X, BB}$  &  $L_{\rm X, BB}/L_{\rm bol}$ &  log $T_{\rm X, BB}$  &  log $R_{\rm X, BB}$  \\
  $(\dot{M}_{\rm Edd})$  & ($L_{\rm Edd}$) & ($^\circ$) &   (erg $\rm s^{-1}$) &   &   (K) &   (cm) &  (erg $\rm s^{-1}$)  &   &  (K) &    (cm)
 \\  [0.3cm] \hline

       %7 & 1 & \makecell[l]{ 10    \\ 30  \\ 50\\70}    & \makecell{ 10    \\ 30  \\ 50\\70}        \\
       \multirow{8}{*}{7} & \multirow{4}{*}{1} & 10 &   40.650 &     0.000 &         4.47 &          12.71  &   43.952 &    0.710 &         5.99 &          11.38  \\  \cline{3-11} 
        &  &   30 &     42.212 &     0.013 &         4.89 &          12.64 &   43.742 &    0.438 &         5.91 &          11.45 \\   \cline{3-11} 
        &  &   50 &   42.480 &     0.024 &         4.67 &        13.48 &   42.475 &    0.024 &         5.72 &          11.85 \\   \cline{3-11} 
        &  &   70   & 43.179 &     0.120 &         4.30 &          14.55 &   38.879 &    0.000 &         5.25 &          13.11 \\    \cline{2-11} 
        & \multirow{4}{*}{7}  & 10 & 42.861 &     0.008 &        5.13 &    12.50 &   44.799 &    0.713 &         5.99 &         11.80 \\ \cline{3-11} 
        &  &  30 &           43.434 &     0.031 &         5.04 &    12.95 &   44.587 &    0.438 &         5.91 &          11.87 \\\cline{3-11} 
        &  & 50 &            43.691 &     0.056 &         4.85 &    13.72 &   43.407 &    0.029 &         5.74 &          12.29 \\\cline{3-11} 
        &  &  70 &            43.175 &     0.017 &         4.30 &   14.55 &   41.422 &    0.000 &         5.35 &          13.55 \\[0.2cm]\hline

        \multirow{8}{*}{12} & \multirow{4}{*}{1} & 10 &  40.989 &     0.001 &         4.55 &      12.73 &   43.914 &    0.651 &         5.98 &      11.38 \\\cline{3-11} 
        &  &  30 &   42.503 &     0.025 &         4.74 &    13.09 &   43.334 &    0.171 &         5.78 &          11.61 \\ \cline{3-11} 
        &  &  50 &   42.684 &     0.038 &         4.54 &   13.84 &   41.451 &    0.002 &         5.57 &          12.05 \\\cline{3-11} 
        &  & 70 &    43.508 &     0.255 &         4.38 &    14.57 &     - &     - &         - &          - \\\cline{2-11} 
        
         & \multirow{4}{*}{12} &  10 & 43.364 &     0.015 &     5.13 &    12.75 &   44.994 &    0.652 &         5.99 &    11.91 \\ \cline{3-11} 
        &  &  30 &    43.921 &     0.055 &         4.89 &    13.50 &   44.413 &    0.171 &         5.78 &         12.15 \\ \cline{3-11} 
        &  &   50 &   44.130 &     0.089 &         4.78 &    14.08 &   42.973 &    0.006 &         5.65 &         12.40 \\ \cline{3-11} 
        &  &  70 &    43.558 &     0.024 &         4.34 &    14.66 &   - &    - &      - &           - \\ [0.2cm]\hline
        
        \multirow{8}{*}{24} & \multirow{4}{*}{1} & 10 &  41.407 &     0.002 &         4.44 &    13.15 &   43.786 &    0.483 &         5.98 &          11.30 \\ \cline{3-11} 
        &  & 30 &     42.994 &     0.078 &         4.53 &   13.75 &   41.452 &    0.002 &         5.40 &          12.17 \\ \cline{3-11} 
        &  &  50 &    43.193 &     0.123 &         4.48 &   14.20 &   - &    - &      - &           - \\ \cline{3-11} 
        &  & 70 &     43.784 &     0.482 &         4.41 &   14.64 &      - &      - &     - &           - \\  \cline{2-11} 
    
        & \multirow{4}{*}{24} & 10 &         43.427 &     0.009 &         4.83 &   13.37 &   45.167 &    0.485 &         5.99 &     11.97 \\  \cline{3-11} 
        &  &  30 &        44.790 &     0.204 &         4.77 &   14.18 &   42.977 &    0.003 &         5.43 &        12.80 \\  \cline{3-11} 
        &  & 50 &        44.401 &     0.083 &         4.76 &    14.26 &   43.026 &    0.004 &         5.50 &          13.22 \\  \cline{3-11} 
        &  &  70 &    44.444 &     0.092 &         4.67 &      14.46 &      - &      - &     - &       -  \\ [0.2cm]\hline \\
\end{tabular}
\label{tab:modelobservable}
\end{table}

\section{Characteristic Timescales}
\label{app:timescale}

\noindent 
There are various physical timescales relevant fo TDEs: debris mass fallback timescale, disk formation/circularization timescale, disk viscous timescale, and photon transport timescale. The longest timescale of all governs the temporal evolution of TDE emissions. 

The photon transport time through optically thick medium is the shorter one between the diffusion timescale and advection timescale. The photon diffusion timescale is calculated as $t_{\rm diff}=\tau_{\rm es} \  R_{\rm es}/c$, where $R_{\rm es}$ is the size of the electron-scattering photosphere along a particular inclination, and $\tau_{\rm es}$ is the electron-scattering optical depth integrated radially from $r=0$ to $r=R_{\rm es}$. 
The advection timescales is calculated as $t_{\rm adv} \approx R_{\rm es}/v_r$, since the photons trapped by the optically-thick gas has a similar speed as the gas. Here the gas radial velocity $v_r$ is averaged over the radial path within $R_{\rm es}$ and weighted by gas density. The values for the $\dot{M}_{\rm acc} = 12 \dot{M}_{\rm Edd}$ disk are given in Table \ref{tab:photontimes}. One can see that for the inclinations considered in this work, photons are preferably advected out by the optically-thick wind. The photon transport time varies from $\sim$0.1 day to a few days depending on the inclination.

The disk viscous timescale can be analytically calculated by $t_{\rm visc}=t_{\rm dyn} \ \alpha^{-1} (H/R)^{-2}$, where $t_{\rm dyn}$ is the orbital timescale of the disk, $\alpha$ is a  
free parameter between 0 and approximately 1, and $H/R$ is the disk thickness \citep{Shakura73}. For our disk parameters, we have 
$t_{\rm visc}\approx 5.44 {\rm \ days} \ (R_{\rm disk}/8500 r_g) \ (M_{\rm BH}/10^6M_\odot) (\alpha/1)^{-1} ((H/R)/0.3)^{-2}$. The viscous time is therefore only a few days. We caution the readers that our simulated disks are MAD which typically have effective $\alpha \gtrsim 1$. The viscous time is potentially longer if the disks do not have such large magnetic fluxes.

The disk formation/circularization timescale induces uncertainty into the evolution of TDEs. Recent simulations show that a large fraction of the debris materials can form a disk within dynamical timescale but the disk still pocesses some moderate eccentricity \citep{Bonnerot21}. As the topic is out of the scope of this paper, we assume here that the disk forms quickly for the calculation of the emission evolution.

Stellar debris typically falls back on timescales of $t_{\rm mb}$, which is the orbital time of the most bound debris \citep{Evans89, Guillochon13, Rossi21review}:
\begin{equation} \label{eq:tmb}
    t_{\rm mb}\approx 41 \ {\rm days} \ \Big( \frac{M_{\rm BH}}{10^6 M_\odot}\Big)^{1/2} \Big( \frac{m_\star}{M_\odot}\Big)^{-1} \Big( \frac{r_\star}{R_\odot}\Big)^{3/2}.
\end{equation}
For our case with $M_{\rm BH} = 10^6 M_\odot$, $t_{\rm mb}$ varies between 25 days and 41 days for $m_\star = (0.1- 10) \  M_\odot$. 

Lastly, assuming that the fallback timescale governs the evolution of TDEs, we calculate the time corresponding to each simulated disk with a further assumption that the fallback rate $\dot{M}_{\rm fb}$ equals the instataneous accretion rate $\dot{M}_{\rm acc}$ plus the wind mass rate $\dot{M}_w$. Here we focus on the post-peak evolution. By setting that the fallback rate peaks at $t_{\rm mb}$, the post-peak time $t_{\rm pp}$ associated with a particular $\dot{M}_{\rm fb}$ can be calculated with:  
\begin{equation} \label{eq:MFB}
   \dot{M}_{\rm fb}\approx 133 \Big( \frac{M_{BH}}{10^6 M_\odot}\Big)^{-3/2} \Big( \frac{m_\star}{M_\odot}\Big)^2 \Big( \frac{r_\star}{R_\odot}\Big)^{-3/2} \Big( \frac{t}{t_{\rm mb}}\Big)^{-5/3} \dot{M}_{\rm Edd}.
\end{equation} 
With the wind mass rate given in Table \ref{tab:quantities}, the three simulated disks correspond to $\dot{M}_{\rm fb}= \dot{M}_{\rm acc} + \dot{M}_w = (7+1.4, 12+4.5, 24+14) \ \dot{M}_{\rm Edd}$. 
The post-peak time obtained with $M_{\rm BH} = 10^6 M_\odot$ and a few different $m_\star$ are shown in Table \ref{tab:postpeaktime}.

\begin{table}[]
\centering
\caption{Photon diffusion and advection timescales  for the simulated disk at $\dot{M}_{\rm acc}=12\dot{M}_{\rm Edd}$}
\begin{tabular}{l|cccc}
\tableline
inclination ($^\circ$) & 10 & 30 & 50 & 70 \\ \hline
 $R_{\rm es} \  (R_g)$ &         896 &         1278 &        5425 &        8377 \\ \hline
 $\tau_{\rm es}$ &         7.0 &         74.0 &         137.2 &         710.2 \\ \hline
      $v_r \ (c)$  &         0.50 &         0.35 &       0.13 &         0.013 \\ \hline
 $t_{\rm adv}$ (day) &         0.07 &         0.13 &        1.06 &         13.6 \\ \hline
 $t_{\rm diff}$ (day) &         0.35 &        5.4 &      42.2 &      337.4 \\ \hline
\end{tabular}
\label{tab:photontimes}
\end{table}     
     
 \begin{table}[]
\centering
\caption{Time  elapsed since peak corresponding to the three simulated TDE disks, assuming $M_{\rm BH} = 10^6 M_\odot$ and various different $m_\star$}
\begin{tabular}{l|cccc}
\tableline  
  $m_\star$ ($M_\odot$) & 0.2 &   0.5 &    1 &  1.5 \\ \hline
  $\dot{M}_{\rm fb, peak} \  (\dot{M}_{\rm Edd})$ & 36.7  &         76.4 &        133.0 &        211.6  \\ \hline
 % $t_{\rm peak}$ & & & \\
 $t_{\rm pp}(\dot{M}_{\rm fb}= (24 + 14) \ \dot{M}_{\rm Edd} )$ (day) &         NA  &    18.6      &      45.9 & 69.7     \\ %\hline
    $t_{\rm pp}(\dot{M}_{\rm fb}= (12 + 4.5) \ \dot{M}_{\rm Edd} )$ (day) &  18.3       &      53.8  &     102.4   &   140.0      \\ %\hline
   $t_{\rm pp}(\dot{M}_{\rm fb}= (7 + 1.4) \ \dot{M}_{\rm Edd} )$ (day) &  42.3    &   98.5    &  174.0     &  229.2      \\ \hline

\end{tabular}
\label{tab:postpeaktime}
\end{table}

\end{document}